%% file: ms.tex
\documentclass[preprint2]{aastex}

\usepackage{epstopdf}
\usepackage{graphics, subfigure}
\usepackage{lscape}
\usepackage{psfig}
\usepackage{multicol}
\usepackage{breqn}

\input{defs.txt}

\def\grb{GRB\,141121A}

\begin{document}

\title{Happy Birthday \swift: Ultra-long GRB\,141121A and its broad-band Afterglow}


\author{
	A.~Cucchiara\altaffilmark{1},
	P. ~Veres\altaffilmark{2},
	A.~Corsi\altaffilmark{3},
	S.~B. Cenko\altaffilmark{4,5},
	D.~A. Perley\altaffilmark{6,7},
	A.~Lien\altaffilmark{8,9},
	F.~E.~Marshall\altaffilmark{4},
	C. Pagani\altaffilmark{10},
	V.~L.~Toy\altaffilmark{11},
	J.~I.~Capone\altaffilmark{11},
	D.~A.~Frail\altaffilmark{12},
	A.~Horesh\altaffilmark{6},
	M.~Modjaz\altaffilmark{13}, 
	N.~R.~Butler\altaffilmark{14},  
	O.~M.~Littlejohns\altaffilmark{14}, 
	A.~M.~Watson\altaffilmark{15}, 
	A.~S.~Kutyrev\altaffilmark{4}, 
	W.~H.~Lee\altaffilmark{15}, 
	M.~G.~Richer\altaffilmark{15}, 
	C.~R.~Klein\altaffilmark{16}, 
	O.~D.~Fox\altaffilmark{16}, 
	J.~X.~Prochaska\altaffilmark{17}, 
	J.~S.~Bloom\altaffilmark{16}, 
	E.~Troja\altaffilmark{4}, 
	E.~Ramirez-Ruiz\altaffilmark{17}, 
	J.~A.~de~Diego\altaffilmark{15}, 
	L.~Georgiev\altaffilmark{15}, 
	J.~Gonz\'{a}lez\altaffilmark{15}, 
	C.~G.~Rom\'{a}n-Z\'{u}\~{n}iga\altaffilmark{15}, 
	N.~Gehrels\altaffilmark{4} and 
	H.~Moseley\altaffilmark{4}
	}
	\email{antonino.cucchiara@nasa.gov}

\altaffiltext{1}{NASA Postdoctoral Program Fellow, Goddard Space Flight Center, Greenbelt, MD 20771, USA}
\altaffiltext{2}{The George Washington University, Department of Physics, 725 21st, NW Washington, DC 20052, USA}
\altaffiltext{3}{Physics Department, Texas Tech University, Box 41051, Lubbock, TX 79409, USA}
\altaffiltext{4}{Astrophysics Science Division, NASA Goddard Space Flight Center, MC 661, Greenbelt, MD 20771, USA}
\altaffiltext{5}{Joint Space-Science Institute, University of Maryland, College Park, MD 20742, USA}
\altaffiltext{6}{Department of Astronomy, California Institute of Technology, MC 249-17, 1200 East California Blvd., Pasadena, CA 91125, USA}
\altaffiltext{7}{Dark Cosmology Centre, Niels Bohr Institute, University of Copenhagen, Juliane Maries Vej 30, DK-2100 Copenhagen, Denmark}
\altaffiltext{8}{Center for Research and Exploration in Space Science and Technology (CRESST) and NASA Goddard Space Flight Center, Greenbelt, MD 20771, USA}
\altaffiltext{9}{Department of Physics, University of Maryland, Baltimore County, 1000 Hilltop Circle, Baltimore, MD 21250, USA}

\altaffiltext{10}{Department of Physics and Astronomy, University of Leicester, University Road, Leicester LE1 7RH, UK}

\altaffiltext{11}{Department of Astronomy, University of Maryland, College Park, MD 20742, USA}

\altaffiltext{12}{National Radio Astronomy Observatory P.O. Box 0. Socorro, NM} 
\altaffiltext{13}{Center for Cosmology and Particle Physics, New York University, 4 Washington Place, New York, NY 10003, USA}

\altaffiltext{14}{School of Earth \& Space Exploration, Arizona State University,    AZ 85287, USA}
\altaffiltext{15}{Instituto de Astronom\'{i}a, Universidad Nacional Aut\'{o}noma de M\'{e}xico, Apartado Postal 70-264, 04510 M\'{e}xico, D. F., M\'{e}xico}
\altaffiltext{16}{Astronomy Department, University of California, Berkeley,
    CA 94720-7450, USA}
\altaffiltext{17}{Department of Astronomy and Astrophysics, UCO\/Lick Observatory, University of California, 1156 High Street, Santa Cruz, CA 95064, USA}

\begin{abstract}
We present our extensive observational campaign on the \swift-discovered \grb, almost ten years after its launch. Our observations covers radio through X-rays, and extends for more than 30 days after discovery. The prompt phase of \grb\, lasted  1410\,s and, at the derived redshift of $z=1.469$,  the isotropic energy is  $E_{\gamma,\mathrm{iso}}=8.0\times 10^{52}$ erg. Due to the long prompt duration, \grb\ falls into the recently discovered class of UL-GRBs. Peculiar features of this burst are a flat early-time optical light curve and a radio-to-\xray\ rebrightening around 3 days after the burst. The latter is followed by a steep optical-to-X-ray decay and a much shallower radio fading. We analyze \grb\ in the context of the standard \emph{forward-reverse} shock (FS,RS) scenario and we disentangle the FS and RS contributions.
Finally, we comment on the puzzling early-time ($t\lesssim 3\,$d) behavior of \grb, and suggest that its interpretation may require a two-component jet model. Overall, our analysis confirms that the class of UL-GRBs represents our best  opportunity to firmly establish the prominent emission mechanisms in action during powerful GRB explosions, and future missions (like SVOM, XTiDE, or ISS-Lobster) will provide many more of such objects.

\end{abstract}


\section{Introduction }
\label{sec:intro}
Gamma-ray Bursts (GRBs) have been studied for more than four decades since their discovery. The \swift\ satellite \citep{Gehrels:2004fj} 
has revolutionized our knowledge of their low-energy and long-lasting emission, the afterglow. In fact, this satellite's fast-slewing capability and the
\xray/Optical instruments onboard provide prompt (within minutes) and very accurate ($\sim$ few arcseconds) GRB localization to ground-based observers: since ten years from its launch, on 20 November, 2004, every year \swift\ has dispensed exciting discoveries opening new windows into ``time-domain'' astronomy \citep[see e.g.][]{Bloom:2011aa,tfl09,Tanvir:2013aa,Gal-Yam:2006yq,Soderberg:2006vn,Berger:2014aa}. Moreover, \swift\ has discovered more than 900 GRBs, the vast majority of which belong to the \emph{long} class, with a duration of the gamma-ray emission, $T_{90}$, larger than two seconds \citep{Kouveliotou:1993aa}. 
Long-duration GRBs are associated with the core-collapse of massive stars \citep[e.g. ][]{Woosley:2006aa}, although the precise nature of their progenitors is still being investigated. The study of the long-lasting afterglow in the temporal and spectral domains enables the characterization of the emission mechanism, the geometry of the ejecta, and the structure of the progenitor surrounding environment \citep{Sari:1998kl}. 

In the \textit{fireball model} \citep{Meszaros:1993qy}, afterglow emission arises from a forward shock (FS) impacting on the external medium, and early emission from a reverse shock (RS) is also expected. Typically, RS observables are  prompt optical and radio flashes \citep[see GRB\,990123, ][]{Akerlof:1999aa,Kulkarni:1999aa,Meszaros:1999aa,Sari:1999aa,Kobayashi:2003aa}. However, despite many years of research and the increased number of rapid response observations from robotic facilities, RS signatures have been detected in surprisingly few cases \citep{Melandri:2008aa,Cucchiara:2011aa,van-der-Horst:2014aa,Vestrand:2014aa,Perley:2014aa,Gendre:2012aa,Laskar:2013aa}. 

Disentangling the RS emission from other possibilities (such as refreshed shock emission or double-jet hypothesis) which mimic the observed temporal and spectral behavior is 
a challenging task, which requires ample datasets in the temporal-spectral regimes.
In the radio, only recently, thanks to the upgraded Karl G. Jansky Very Large Array \citep[VLA\footnote{The National Radio Astronomy Observatory is a facility of the National Science Foundation operated under cooperative agreement by Associated Universities, Inc.},][]{Perley:2009aa}, we have been able to reach the sensitivity required to search for RS in GRB afterglows using multi-wavelength datasets spanning the 1-100\,GHz range \citep{Veres:2014aa,Laskar:2013aa,Perley:2014aa}. 
While FS provides constraints on the circumburst medium, the RS radio-to-optical emission provides a unique tool to investigate the properties of the jetted emitting region (e.g. the initial Lorentz factor $\Gamma$ and the magnetization of the ejecta).

The recent identification of \emph{ultra-long} GRBs \citep[UL-GRB,][]{Virgili:2013aa,Levan:2014aa,Evans:2014aa} has opened a new opportunity to study these explosive phenomena. The exact emission mechanism and progenitor of UL-GRBs is still debated (their prompt emission usually lasts $\gtrsim 1000$\,s). If UL-GRBs share with long GRBs similar progenitors, but occur in a low-density medium \citep[as recently proposed by ][but see also \citealt{Stratta:2013aa} and reference therein]{Evans:2014aa,Piro:2014aa}, the acquisition of radio data is crucial because it enables the characterization of the circumburst density, thus providing a test for this scenario. Furthermore, if UL-GRBs are associated with low-density environments, then the deceleration time of the fireball (at which point the FS afterglow emission starts) would be delayed. In the fireball model (assuming a thin shell case), the deceleration time also marks the peak of the RS emission \citep{Sari:1998kl}, and UL-GRBs may help us find RSs at much later times (Section~\ref{sec:discussion}).

Here, we present our multiband observations of the UL-GRB 141121A, detected by \swift\ almost exactly ten years after its launch. Using our approved radio programs\footnote{VLA/14A-430, PI: A. Corsi; VLA/14B-490, PI: A. Corsi} and the Reionization and Transients Infrared telescope \citep[RATIR,][]{Butler:2012aa}\footnote{http://www.ratir.org}, we were able to follow the afterglow behavior of this burst starting only a few hours after the discovery, until one month later. The paper is organized as follows: in Section \ref{sec:obs} we present our rich dataset; in Section \ref{sec:analysis} we discuss our temporal and spectral analysis in light of the FS-RS scenario, while in Section \ref{sec:discussion} we investigate the implication of our model and alternative possibilities. Finally, Section \ref{sec:conclusions} summarizes our findings.
 
Throughout the paper we approximate the afterglow brightness as composed by a series of power-law 
segments ($F(t,\nu)\propto \nu^{-\beta}t^{-\alpha}$). We will use the standard cosmological parameters, $H_0 = 70$\,km s$^{-1}$\,Mpc$^{-1}$, $\Omega_m = 0.27$, and $\Omega_{\Lambda} = 0.73$.


\section{Observations}
\label{sec:obs}
\subsection{Space-based Observations}
\grb\ was discovered by the Burst Alert Telescope \citep[BAT,][]{Lien:2014aa,Barthelmy:2005lr}
on-board \swift\ at 03:50:43 UTC ($T_{BAT}$) on 2014 November 21. 
The time-averaged spectrum from $T_{BAT}+110.3$ to $T_{BAT}$+663.0\,s is best fit by a simple
power-law model \citep[Eq. 1 in ][]{Sakamoto:2011aa} with photon index  $1.74\pm0.13$. The fluence in the 15-150\,keV band is $F_{\gamma}=(4.3 \pm 0.4) 
\times 10^{-6}$\,erg\,cm$^{-2}$. All quoted errors are at the 90\% confidence level.

 The burst was also detected by the Monitor of All-sky X-ray Image Gas Slit Camera (MAXI/GSC) 
 instrument on-board the International Space Station almost 6 minutes before the BAT trigger \citep{Honda:2014aa}.
 This early emission was also seen by the Konus-Wind  \citep{Golenetskii:2014aa}: significant flux excess was detected in the 20 keV to 10 MeV energy range with a fluence of $F_{\gamma}=8 \times 10^{-6}$\,erg\,cm$^{-2}$. Konus-Wind also observed \grb\,
during the BAT trigger, putting this GRB in the class of UL-GRBs \citep[see Section \ref{sec:analysis},][]{Levan:2014aa}. 
For the rest of the paper we consider the Konus-Wind detection as the starting time of the GRB, $T_0=T_{BAT}-860$\,s, and therefore the overall duration of \grb\ is $T=1410$\,s.
At a redshift of $z=1.469$ (Section \ref{sec:spec}), we estimate an isotropically emitted energy of $E_{\rm iso}=8.0\times 10^{52}$\,erg within the Konus-Wind energy range.

The \swift\ X-ray Telescope \citep[XRT,][]{Burrows:2005fk} started observations of \grb\ 355\,s after the BAT trigger, collecting data in Windowed Timing (WT) settling mode while the spacecraft was slewing to the burst location. The X-ray afterglow was localized in an image taken 362\,s after the BAT trigger; the astrometrically corrected X-ray position \citep{ebp07}, derived using the XRT-UVOT alignment and matching UVOT  \citep[UltraViolet and Optical Telescope,][]{Roming:2005qy} field sources to the USNO-B1 catalogue is $\alpha = 08^{\rm h}10^{\rm m}40^{\rm s}.67$, $\delta = +22^\circ13'02\farcs7$ (equinox 2000.0) with an estimated uncertainty
of $1\farcs5$ (radius, 90\% confidence including systematic error). 
Settled observations in WT mode started at $T_{\rm BAT}+369$\,s until $T_{\rm BAT}+3.9$\,ks, and data in Photon Counting (PC) mode was acquired 
from $T_{\rm BAT}+5.5$\,ks to $T_{\rm BAT}+1.48$\,Ms. The total exposure time was 118.6\,ks.
The XRT event files were processed using the standard pipeline software ({\scshape xrtpipeline} v0.13.1), applying the default filtering and screening criteria ({\scshape HEASOFT 6.16}),
using the latest CALDB 4.4 files released in September 2014.

\begin{figure*}[t!]
\epsscale{1.0}
\includegraphics[scale=0.75,angle=0]{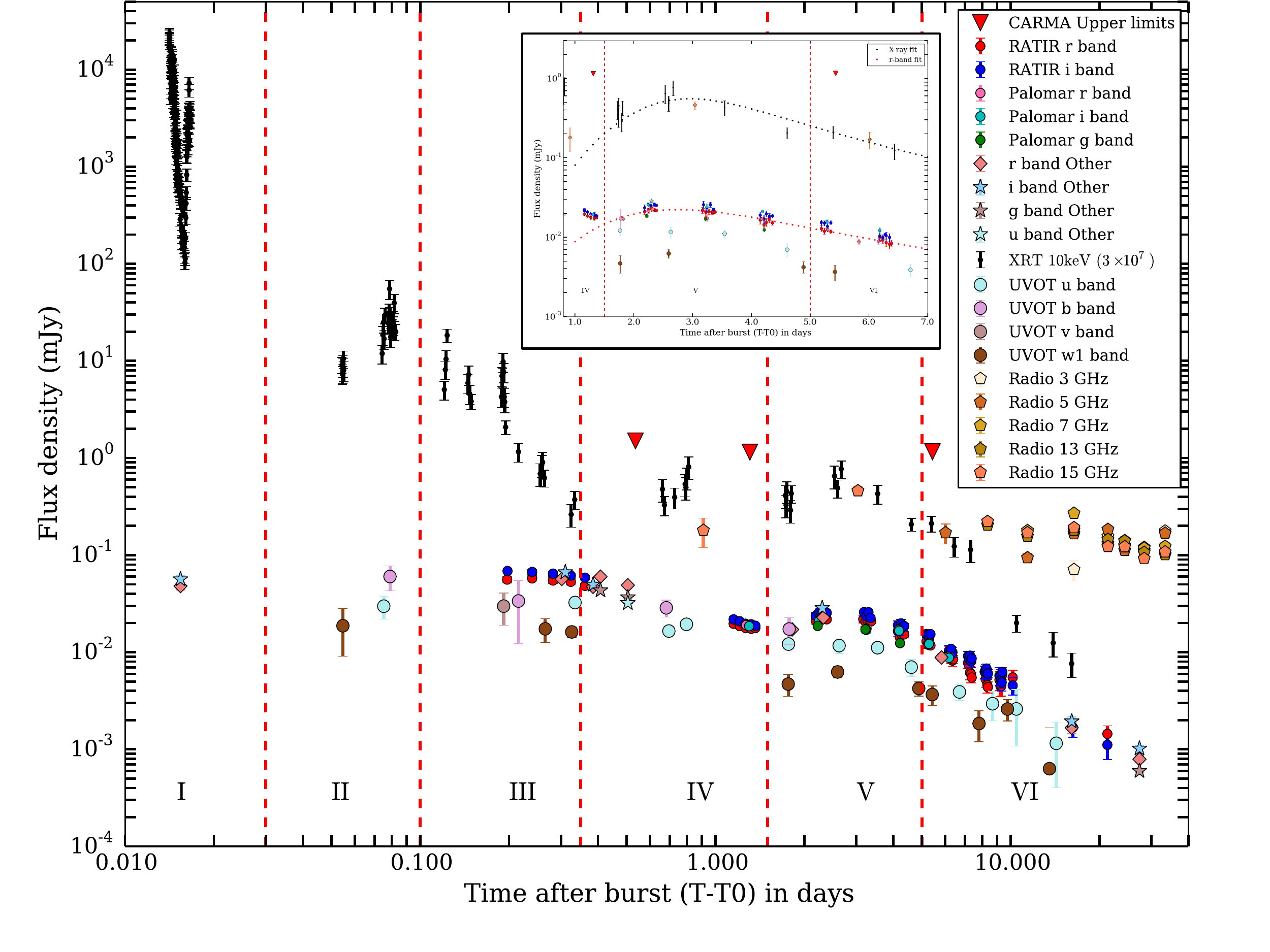}
\caption{\footnotesize{\grb\ light curve: we divided the light curve in 6 regions of interest. An achromatic peak 
is evident at $t\approx 3$\,d. In the inset we zoom in this region and over plot to the optical \rp\ and \xray\ data the best fit
for the broken power law (see Table~\ref{tab:specfit}).}
}
\label{fig:lctotal}
\end{figure*}

The X-ray light curve of the afterglow presented in Figure~\ref{fig:lctotal} was obtained from the Burst Analyser repository\footnote{\url{http://www.swift.ac.uk/burst_analyser/00619182/}}, maintained by the XRT team at the University of Leicester. The light curve, in units of mJy at 10\,keV, was extracted using the methods described in \cite{ebp09,Evans:2010aa}.

Time resolved X-ray spectra of the afterglow in the energy range 0.3--10\,keV were extracted for six regions (see later sections). Only grade 0 to 12 events were selected for PC mode data, binning the data in energy with 1 count per bin. {\scshape xspec} v12.8.2 was used for the spectral analysis. 
An absorbed power-law model was chosen to fit each spectrum, fixing the Galactic absorption to the value in the direction of the GRB of $N_H=4.28\times 10^{20}\, {\rm cm}^{-2}$, as calculated from \citet{Willingale:2013aa} and using the TBabs and ZTBabs absorption models at the GRB redshift of $z=1.469$, with the \cite{Wilms:2000aa} abundances. The X-ray fluxes used in the SED analysis, in units of erg\,cm$^{-2}$\,s$^{-1}$, were derived from the best fit results of the spectral modeling for the six selected time intervals (see Table~\ref{tab:specfit}).

The UVOT began settled observations of\grb\ 371\,s after
the BAT trigger. Initially exposures were taken with all 6 lenticular
filters plus the open (white) filter, but after about $T_{BAT} + 25$\,ks,
almost all the exposures used either the \emph{u} or \emph{uvw1} filters,
with central wavelengths of 346 nm and 260 nm respectively.
Aperture photometry as described by \citealt{Poole:2008aa} was
carried out for each exposure using the standard {\scshape HEASOFT 6.16}
tools 
and the latest UVOT calibration \citep{Breeveld:2010aa,Breeveld:2011aa}.
A 3\arcsec\ radius aperture was centered on the position
determined from the 6 UVOT exposures with the best detections
of the afterglow. The measured count rates were corrected
for extinction in the Milky Way using the compilation from \citet{Schlafly:2011aa} and then converted
to fluxes using a standard GRB spectrum \citep[Table 10 in][]{Poole:2008aa}.

\subsection{Optical}
RATIR started observing \grb\, four hours after the burst 
and continued monitoring its optical behavior until 21\,d post burst, when the afterglow fell below the detection limit of the instrument. The optical camera provided \rp 
and \ip\ observations via a usual sequence consisting of a series of optical frames with exposure times of 80\,s each which are reduced in real-time 
using an automatic pipeline \citep[see][ for more details]{Littlejohns:2014aa}. Multiple exposures were combined in order to increase the signal-to-noise
ratio and aperture photometry was performed at the GRB location. Magnitudes were calibrated using nearby point sources from Sloan Digital Sky Survey (SDSS; \citealt{Ahn:2014aa}).

We imaged the location of GRB\,141121A with the robotic Palomar
60-inch telescope (P60; \citealt{Cenko:2006aa}) beginning at 9:53\,UT on
2014 November 22. Observations were obtained in the $g^{\prime}$, $r^{\prime}$,
and $i^{\prime}$ filters and continued through 2014 November 27.  All
data were processed using a custom IRAF\footnote{IRAF is distributed by 
the National Optical Astronomy Observatory, which is operated by the 
Association of Universities for Research in Astronomy (AURA) under 
cooperative agreement with the National Science Foundation.} pipeline.
Individual exposures were aligned with respect to astrometry from the SDSS 
using SCAMP \citep{Bertin:2006aa} and stacked with SWarp \citep{Bertin:2002aa}. 
We measured aperture photometry on the afterglow of GRB\,141121A and used nearby 
point sources from the SDSS for photometric calibration. 
The resulting measurements are reported in Table~\ref{tab:data}.


Finally, further observations were carried out by the Discovery Channel Telescope equipped with the Large Monolithic Imager (LMI\footnote{\url{http://www2.lowell.edu/rsch/LMI/LMI.html}}) and 
the Keck I telescope equipped with the Low-resolution Imaging Spectrometer \citep[LRIS, ][]{Oke:1995aa}.
For LMI, we acquired a series of 2 minutes exposures in \gp,\rp\,\ip, and \zp\ filters and performed bias subtraction, flat-fielding correction, and cosmic ray removal using our customized pipeline \citep{Toy:2014aa}. 
A log of all the optical observations is presented in Table~\ref{tab:data}, after correcting for galactic extinction, assuming $E(B-V)=0.05$ \citep{Schlafly:2011aa}. 

\subsection{Radio}
VLA data were reduced and imaged using the Common Astronomy Software
Applications (CASA) package. Specifically, the calibration was performed using
the VLA calibration pipeline V4.2.2. After running the pipeline, we inspected
the data (calibrators and target source) and applied further flagging when
needed. 3C286 was used as flux calibrator. J0830+2410, J0823+2223, and
J0802+1809 were used as phase calibrators. The VLA measurement errors are a
combination of the rms map error, which measures the contribution of small
unresolved fluctuations in the background emission and random map fluctuations
due to receiver noise, and a basic fractional error (here estimated to be
$\approx 5\%$) which accounts for inaccuracies of the flux density calibration.
Theses errors were added in quadrature and total errors are reported in
Table\,\ref{tab:data}. 

We also observed \grb\, using the Combined Array for Research in
Millimeter-wave Astronomy (CARMA\footnote{\url{https://www.mmarray.org/}}) on three occasions between 2014-11-21 UT and
2014-11-26 UT.  Observations were conducted in single-polarization mode with
the 3\,mm receivers tuned to a frequency of 93\,{\rm GHz}, interleaved with
observations of a nearby gain-calibrator, as well as observations of 3C84 for
flux calibration and 0854+201 for bandpass calibration.  Data were reduced
using the Multichannel Image Reconstruction Image Analysis Display (MIRIAD) tool; 
none of the three epochs resulted in a significant detection of
the afterglow. A summary of our upper limits is given in Table~\ref{tab:data}.

\subsection{Spectroscopy}
\label{sec:spec}
\begin{figure*}[t!]
\epsscale{1.0}
\includegraphics[scale=0.70,angle=0]{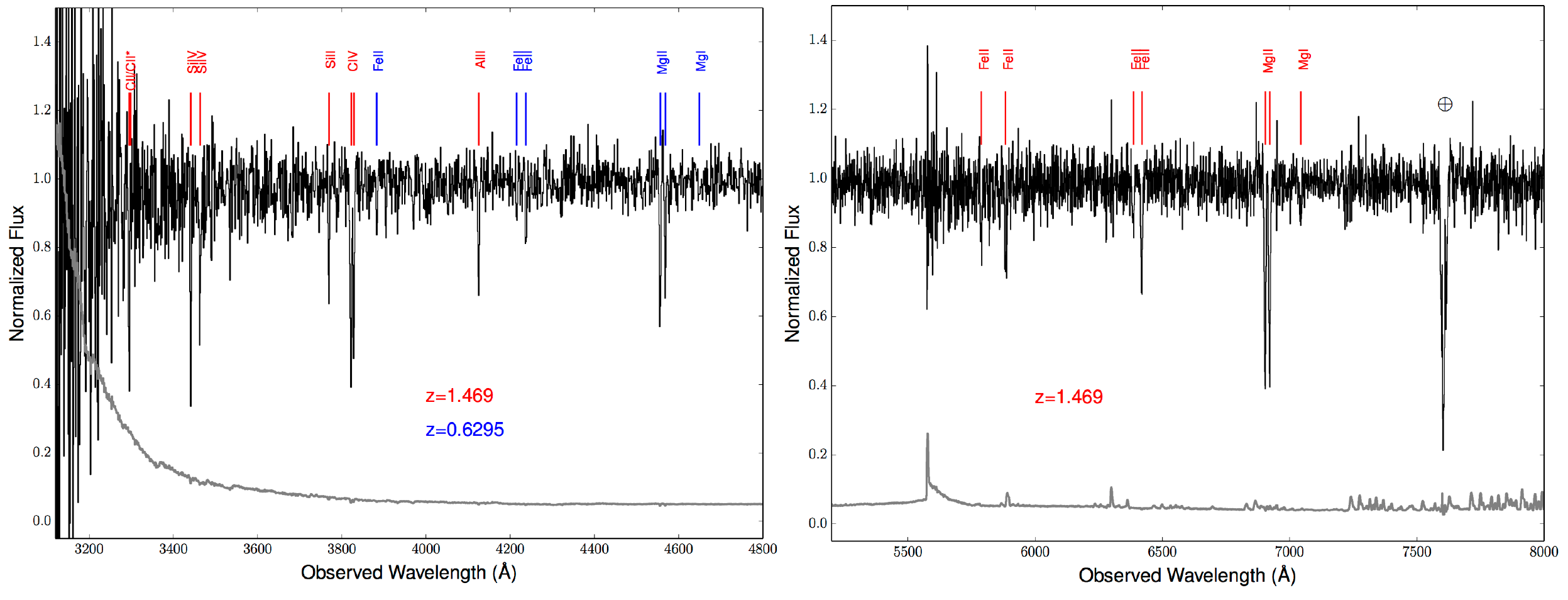}
\caption{\footnotesize{The two panels (separated for simplicity) present two part of the Keck/LRIS spectrum where strong absorption lines belonging to the GRB hosts ($z_1=1.469$) and to an intervening system at $z_2=0.6295$ appear. We also plot in gray the 1-$\sigma$ error array.}}
\label{fig:spec}
\end{figure*}

We acquired spectroscopy of the afterglow of \grb\ using the Low-Resolution
Imaging Spectrograph (LRIS) mounted on the Keck I telescope 
between 11:10:38\,UT and 11:16:18\,UT.  Observations
were taken using the 600/4000 grism on the blue side and 400/8500 grating on
the red side, providing continuous wavelength coverage between 3116--10264 \AA.
Data were reduced in IDL using the LRIS Automated Pipeline
(LPipe\footnote{http://www.astro.caltech.edu/\textasciitilde
dperley/programs/lpipe.html}), with the flux calibration established via a
separate observation of the flux standard BD+28.  The spectrum (Figure
\ref{fig:spec}) presents several absorption features, including
\MgII\ doublet (2796,2803\AA), \FeII\ 2600 and \FeII\ 2586 and \FeII\ 2344 all
at the same redshift of $z=1.4690$. No \Lya\ line is identified down to the
bluer observed wavelengths, providing a stringent upper limit on the GRB
redshift of $z<1.56$.  We also identify an intervening system at $z=0.6295$,
based on \FeII\ and \MgII\ doublet identification.

\subsection{GCN}
We complement our data with results obtained by other observatories and
published in the GRB Coordinates Network \citep[GCN, ][]{Barthelmy:1995lr}. In
particular, we use GCN data that complement our light curve observations.  For
simplicity and to avoid possible cross-calibration issues we used only data
obtained in \rp\ and \ip\ filters (see Table~\ref{tab:data} for the relevant
references).


\section{Analysis}
\label{sec:analysis}
We present in Figure~\ref{fig:lctotal} the radio to X-ray light curve of \grb,
and based on the different temporal and spectral behaviors we decided to divide it
in six different intervals (I to VI), in order to better study the emission mechanisms in 
action at each interval.
In the standard FS-RS scenario, the afterglow emission
is due to synchrotron radiation of shock-accelerated electrons, and we expect
the observed spectrum across a large frequency range to be represented by a
series of joined power-laws with breaks at characteristic frequencies
\citep{Meszaros:1997dk,Sari:1998kl, Granot:2002aa}: a self-absorption frequency
($\nu_{a}$), an injection frequency which identifies the peak of the
synchrotron emission ($\nu_m$), and the cooling frequency ($\nu_c$).  The
spectral indices ($\beta$) are related to the intrinsic shape of the electron energy
distribution (for which a power-law of index $p$ is assumed) and, for a given
circumburst medium (ISM or wind, for example), can be related to the temporal
indices ($\alpha$) by well-known closure relations \citep[e.g., ][]{Racusin:2009aa}. We
report our results for the spectral and temporal indices of \grb\ in
Table~\ref{tab:specfit}. The spectral and temporal behavior of \grb\ in regions
I to VI can be summarized as follows:

\begin{itemize}
\item At $T\lesssim0.1$\,d  (regions I and II) the X-ray afterglow shows large
flaring activity.  The GRB was detected only by the GROND instrument (two hours
post-burst) and by UVOT at a flux level ($F_{\rm Opt}=48\mu{\rm Jy}$) which is
similar for both regions, suggesting minimal variability.

\item In region III ($0.1\,{\rm d} \lesssim T\lesssim0.35$\,d) the X-ray
afterglow behaves similarly to the so-called ``steep decay phase''
\citep{Zhang:2006aa} observed in other GRBs, with a steep temporal slope
($\alpha_{\rm III,X}=3.1 \pm 0.1$, with $\chi^2=7.2$ and $d.o.f=9$) and a typical spectral index
($\beta_X=0.92\pm0.17$), despite a hint of flare is present at/around $\sim 2$\,d.  On the other hand, the optical light curve shows a
much shallower decay ($\alpha_{\rm III,Opt}=0.15 \pm 0.11$, $\chi^2=0.2$ and $d.o.f=2$) and a similar spectral index $\beta_{\rm Opt}=0.87\pm0.02$.  This suggests a different origin for
the X-ray and optical emission during this time interval. We interpret the
\xray\ behavior as a combination of high-latitude emission \citep{Kumar:2000aa}
superimposed to some contribution from the original prompt phase, as  seen in
many other bursts  \citep[see for example][and Section~\ref{sec:discussion}
later on]{Nousek:2006aa,Racusin:2009aa,Genet:2009aa}. The optical behavior
during this time interval (and in regions I and II) is puzzling, and we will
discuss possible intepretations in the next sections.

\item In region IV ($0.35\,{\rm d}  \lesssim T\lesssim1.5$\,d) the optical/UV
light curve can be fitted by a single power-law with temporal decay index
$\alpha_{\rm IV,Opt}=0.84 \pm 0.11$ (standard for afterglow-dominated
emission), and spectral index $\beta=0.36\pm0.21$ (harder than a typical
afterglow index). In the \xray, instead, we see a constant flux, similar to the
canonical ``plateau'' phase \citep{Racusin:2009aa}, but there is a hint of a
possible flare around $T\approx 0.8$\,d right at the end of the \swift\ orbit.

\item During region V ($1.5$\,d $\lesssim T\lesssim5$\,d), at $\approx 3$\,d after the burst, 
we observe a peak in both the X-ray and optical bands. AMI observations at 14.5\,GHz also hint 
to the presence of a peak around the same time ($\approx 3$\,d).
We fit the optical and X-ray light curves with a smoothly broken
power-law \citep{Beuermann:1999aa}: 
{\footnotesize $F_\nu(t)=F_0 \left[ (t/t_{\rm break})^{s \alpha_{\rm rise}}+ 
(t/t_{\rm break})^{s \alpha_{\rm decay}}  \right]^{-1/s}$}, 
 where we set the roundness parameter to $s=1$.
The broken power-law in the X-ray has
the following parameters: $\alpha^{\rm rise}_{\rm X}=-2.33 \pm 0.88 $, $\alpha^{\rm
decay}_{\rm X} = 2.86 \pm 0.21 $, $t_{\rm peak}^{\rm X}=3.06 \pm 0.71$\,d ($\chi^2=7.4$ and $d.o.f=8$), while in the
optical: $\alpha^{\rm rise}_{\rm Opt}=-1.77 \pm 0.77 $, $\alpha^{\rm decay}_{\rm Opt} = 1.84
\pm 0.17 $, $t_{\rm peak}^{\rm Opt}=3.53 \pm 0.27$\,d (with a $\chi^2=13.4$ and $d.o.f=29$). 
A single power law fit does not provide a good representation of such data with
a $\chi^2=154$ and $d.o.f=28$ (optical) and $\chi^2=21$ and $d.o.f=8$ (\xray).
The optical ($\beta_{\rm Opt}=0.78\pm0.28$) 
and X-ray ($\beta_X=0.67\pm0.23$) spectral indices show no strong evidence for a spectral break between the two bands within the errors.

\item In Region VI (T $\gtrsim 5$\,d) we observe a consistent decay, $\alpha_{\rm VI,Opt}=2.06 \pm
0.40$ ($\alpha_{\rm VI,X}=2.14 \pm 0.34$), in both the \xray\ and optical
bands ($\chi^2=24$ and $d.o.f=21$ and $\chi^2=3.7$ and $d.o.f=4$ for the optical and \xray\
respectively), and the spectral indices are also consistent within the errors.  The
radio afterglow at 15\,GHz has been monitored since 3 days post bursts and it decays as $\alpha_{15 GHz}=0.57\pm0.10$ until 11 days
($\chi^2=2.9$ and $d.o.f=3$). Later observations in the 3-15\,GHz range show a flat
temporal decay and a soft-to-hard evolution, suggesting a peak sweeping through all the
radio frequencies (3-15\,GHz; see Table~\ref{tab:specfit} and Section
\ref{sec:discussion}).

\end{itemize}

\section{Discussion}
\label{sec:discussion} 
A re-brightening similar to the one observed for \grb\, in region V 
has been observed also in the case of 
the UL-GRB\,111209A \citep{Yu:2013aa,Stratta:2013aa}. 
Apart from this GRB, only a few other GRBs, not belonging to the UL-GRB class, 
present such peculiar feature, but usually at much earlier times 
\citep[$10^3-10^4$\,s post-burst; e.g. 
GRB\,110213A, GRB\,120326A, GRB\,120404A][]{Cucchiara:2011aa,Guidorzi:2014aa,Melandri:2014aa,Urata:2014aa}.

Overall, \grb\ shares similar characteristics with previously observed UL-GRBs: first,
the duration $T=1410$\,s which could be due, e.g.,  to a prolonged
central engine activity or to a compact central engine embedded in a large progenitor star
\citep[like red supergiant,][]{Quataert:2012aa,
Bromberg:2012aa,Bromberg:2011aa,Woosley:2012aa,Gendre:2013ab}.  
Second, similarly to GRB\,101225A and GRB\,111209A, after the prolonged \xray\ emission, the
light curve rapidly decays (region II). Finally, as pointed out by
\citealt{Levan:2014aa} \citep[see also GRB\,060607A in][]{Ziaeepour:2008aa},
some dips and flaring are sometimes identified after the steep decay phase.
Indeed, in the case of \grb\, we see this kind of behavior during region III.

Our extensive follow-up provides a dataset which is ideal to identify the main
emission mechanisms (FS, RS, or some combination of both) in action during this
burst, and the nature of the surrounding environment (ISM vs.~wind). Hereafter,
we model the FS and RS synchrotron emission as broken power laws, with breaks
at $\nu_a <\nu_m<\nu_c$, with spectral indices $\{-2,-1/3,(p-1)/2,p/2\}$ or
$\{-5/2,-2,(p-1)/2,p/2\}$.

As discussed in the previous Section, \grb\ shows a very complex light curve.
We have identified six different regions with respect to the temporal (and
spectral) properties of its afterglow. In what follows, we start our analysis
from the latest of these regions (region VI, $T\gtrsim 5$\,d), when the
afterglow  of \grb\ seems to settle on a standard power-law decay, and the
flaring/re-brightening episodes observed at earlier times seem to be ceased.
Then, we discuss the earlier epochs in the light of the constraints derived
from region VI.
\subsection{Region VI} \subsubsection{Evidence for a wind medium}
\label{windsec} In region VI, the optical and \xray\ spectral and temporal
indices (see Table~\ref{tab:specfit}) are very similar, suggesting that these
bands are in the same spectral regime of the synchrotron spectrum predicted by
the fireball model. We infer that the most likely scenario is one in which the
emission is dominated by a FS with  characteristic frequencies
$\nu_{m,f}<\nu_{\rm Opt}<\nu_{\rm X}<\nu_{\rm c,f} $.  If we parametrize the
profile of the circumburst density as $n\propto R^{-k}$, we get
\citep[e.g. from][]{Sari:2000aa}: $k=4/[1+1/(2\alpha_{VI}-3\beta_{VI})]= 1.75\pm0.94$,
consistent with a wind environment surrounding the GRB.  In this case, the
temporal index for the spectral regime $\nu_{m,f}<\nu_{\rm Opt}<\nu_{\rm
X}<\nu_{\rm c,f}$ is $\alpha=(1-3p)/4$, from which we estimate $p=2.67\pm 0.08$
for the power-law index of the electron energy distribution. This last result
is consistent with the value of $p$ derived from the optical-to-X-ray spectral
index $(p-1)/2=\beta_{OX}\approx0.84\pm0.02$, which yields $p=2.68\pm 0.05$.

Simple power law fits to the temporal and spectral evolution in this
region can be found in Table 3. 
The model we will introduce in the following sections (Section \ref{sec:phys}) gives a good description of this region. Here, the optical and X-ray measurements are the most straightforward to interpret and they can be explained by a simple FS component. 
Instead, in the radio (in particular at lower frequencies) the RS still dominates. 
\subsubsection{Source size and scintillation}
\label{scintsec}
\begin{figure*}[t!]
\epsscale{1.0}
\includegraphics[scale=0.60,angle=0]{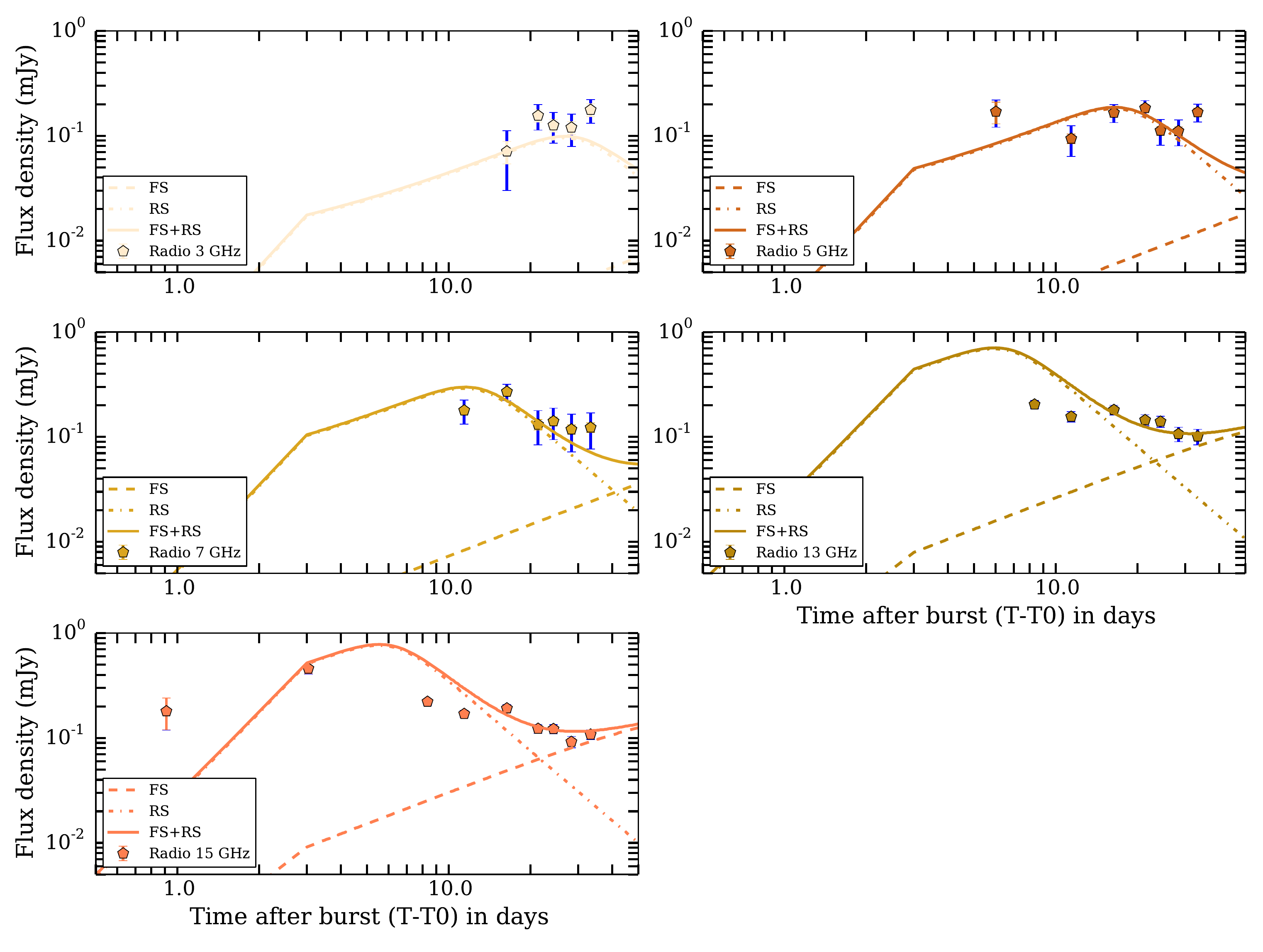}
\caption{\footnotesize{Radio light curves and our model (FS as dashed and RS dotted-dashed) for region VI. We plot with blue errorbars the additional contribution on our error budget from scintillation. While the peak time and the late-time decay  ($\alpha \sim 2.0$) is consistent in all the optical/UV bands, the radio flux presents a much shallower decay ($\alpha_{15GHz}\sim 0.57$). Furthermore there is evidence for a peak sweeping through the 3-7 GHz bands between $\approx 10$\,d and $\approx 30$\,d which we interpret as the passage of $\nu_{SA,f}$. The time evolution of the RS component which on these figures is a power law with two breaks has the following temporal slopes:
$\alpha\approx -2.6,-0.86, 2.2$. The slope of the FS before and after 3 days is
$\alpha\approx -2 {\rm ~and} -1$ respectively.} \label{fig:lclaterad}}
\end{figure*}

As evident from the radio late time light curve (Figure \ref{fig:lclaterad}), the lower-frequency radio data show flux modulations 
that suggest that interstellar scattering and scintillation (ISS) may be important. 
At the location of \grb\ ($l,b\approx200^\circ$,\,$27^{\circ}$), the
characteristic frequency limiting the strong and weak scattering regime is
$\nu_0\approx 12$\,GHz, and the limiting angular size below which (at this
frequency) sources can be considered point sources and exhibit strong
scintillation, is $ \Theta_0\approx2.5\,\mu{\rm as}$ \citep{Frail:2000aa}. 

In the weak scattering regime (in our case, the 14.5\,GHz observations) the predicted 
modulation index can be calculated from \citep{Walker98scint,Walker01scint}:
\begin{equation}
m_{\nu}=(\nu/\nu_0)^{17/12} (\Theta_{\rm source}/\Theta_F)^{-7/6},
\label{modulation:pred1}
\end{equation}
where $\Theta_F=\Theta_0 (\nu/\nu_0)^{-1/2}$. In the strong scattering regime, the predicted 
modulation index is: 
\begin{equation}
m_{\nu}=(\nu/\nu_0)^{17/30} (\Theta_{\rm source}/\Theta_F)^{-7/6},
\label{modulation:pred2}
\end{equation}
with $\Theta_F=\Theta_0 (\nu_0/\nu)^{11/5}$.

From the data at a given frequency, we estimate the observed modulation index 
as in \citep[e.g.][]{Cenko+13orphan,Corsi+14sn} :
\begin{equation}
m_{\nu}= \frac{\sqrt{\langle(F_{\nu}-F_{\nu,pred})^2\rangle -\langle
\sigma_{F_\nu}^2\rangle}}{\langle F_\nu\rangle}
\label{modulation:measured}
\end{equation}
where as predicted flux, $F_{\nu,pred}$, we take a simple power-law fit for
every radio band;  $\sigma_{F_\nu}$ are the measurement errors; and
$\langle\dots\rangle$ denotes the average over time.  From our VLA
observations, we get $m_{3\,{\rm GHz}}\approx 0.3$, $m_{5\,{\rm GHz}}\approx
0.2$, $m_{7\,{\rm GHz}}\approx 0.3$, $m_{13\,{\rm GHz}}\approx 0.1$,
$m_{15\,{\rm GHz}}\approx 0.05$. In Figure \ref{fig:lclaterad} we show in blue
larger error-bars that account for ISS effects.

Using the observed modulation indices (Equation \ref{modulation:measured}) and
comparing them with the predicted ones (Equations \ref{modulation:pred1} and
\ref{modulation:pred2}), we can constrain the apparent size of the emitting
region at $\approx 20-25$\,d since the burst. The most stringent constraint is
derived from the lower frequency observations with the largest modulation
indices. The $3$ GHz observation occurs in the strong scattering regime and so we obtain
$\Theta_{\rm source}$(20\,d)$\approx 76\,\mu{\rm as}$.

We can compare this constraint on the size of the emitting region with the size
predicted by the fireball model for a jet expanding in a wind environment
\citep{Taylor:2004aa}:

\begin{equation} 
\label{eq:latsize}
\Theta=2R_{\perp}/D_A \approx 92 \mu {\rm as}~({E_{54}}/A_{\star,-2})^{1/4} (t/20\,{\rm d})^{3/4}. 
\end{equation}

Here, we have expressed the medium density as $n=AR^{-2}$\,cm$^{-3}$, with
$A=3\times10^{35}A_{\star}$\,cm$^{-1}$.  Thus, if the modulation we observe at
the lowest radio frequencies is indeed due to ISS, then $A_{\star}\approx
2.1\times 10^{-2} E_{54}$. This density parameter is quite close to the one
derived from modeling in Section \ref{sec:phys}.

Finally, because ISS affects more the lower radio frequencies than the higher
ones, its effects need to be taken into account when estimating the radio
spectral indices. To this end, we compare the spectral indices reported in
Table \ref{fig:lclaterad} with the ones we obtain from the best fit power-law
model that we used to measure the observed modulation indices.  At 11\,d,  the
power-law fit gives us $\beta_{radio,pl}\approx-1.5$ (to be compared with the
actual value derived from the data of $\beta_{radio}=-1.64\pm0.32$),  at 16\,d
$\beta_{radio,pl}\approx-0.07$ (to be compared with the actual value derived
from the data of $\beta_{radio}=-1.78\pm0.54$), and at 21\,d
$\beta_{radio,pl}\approx-0.1$ (to be compared with the actual value derived
from the data of $\beta_{radio}=0.18\pm0.07$). Thus, after correcting for ISS
effects, the soft-to-hard evolution observed in the radio band at late times
becomes even more evident, supporting the hypothesis of a spectral break
passing in band.

\subsection{Region V}
\subsubsection{Deceleration time and initial Lorentz factor}
In the fireball model, the afterglow ``starts'' at the deceleration time, 
which is related to the location where the jet sweeps up a fraction $1/\Gamma$ of its mass in 
interstellar material. In a wind case (which, as we have seen in Section \ref{windsec}, 
is the most relevant for \grb), the observed deceleration time is \citep{Zou:2005aa}:
\begin{equation}
t_{\rm dec} = \frac{E (1+z)}{8 \pi A m_p c^3 \Gamma_0^4},
\label{eq:tdec}
\end{equation}  
where $\Gamma_0$ is the Lorentz factor at the deceleration.  Because the
power-law behavior  observed in region VI extends backwards in time to $\approx
3$\,d, we derive $t_{\rm dec}\lesssim 3 $\,d.  Assuming $A_{\star}\approx 0.05$
(as derived in Section \ref{sec:phys}), this implies $\Gamma_0 \gtrsim 27
E_{54}^{1/4} (A_\star/0.05)^{-1/4}$. This matches with the value of $\Gamma_0$
found from modeling in Section \ref{sec:phys}.

\subsubsection{The peak at 3\,d}
\label{sec:3day}
At $\approx 3$\,d, in Region V, a peak (or rebrightening) is observed in the
optical and X-ray light curves of \grb. Because this peak appears to be
achromatic (it is observed in both the optical and X-ray, and there are hints
of a peak at 15\,GHz as well), we consider two scenarios: (i) the peak is
marking the deceleration time of the jet whose emission explains region VI
data; (ii) this last jet is initially off-axis, and its emission enters our
line of sight between 1\,d and 3\,d post-burst, at which time it peaks at all
frequencies.

\begin{figure*}[t!]
\epsscale{1.0}
\includegraphics[scale=0.60,angle=0]{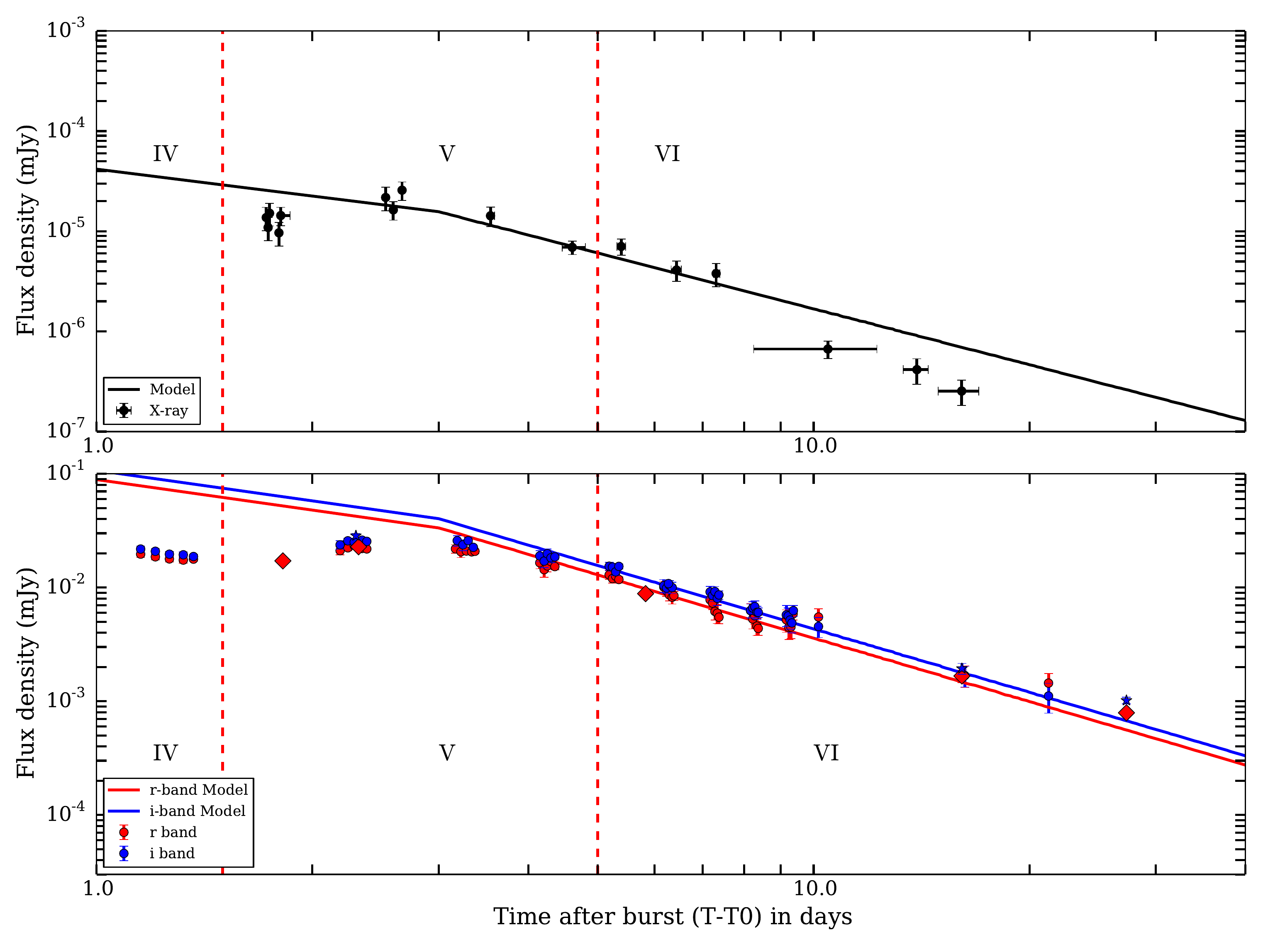}
\caption{\footnotesize{\xray\ (top) and Optical (bottom) late-time light curves (including all the data available in the \rp\ and
\ip\ bands). 
Our model is shown as solid line.
The model (see Section~\ref{sec:discussion}) describes appropriately the data around and after 3\,d.}
}
\label{fig:lclate}
\end{figure*}

In Figure \ref{fig:lclate}, we show our extrapolation of the model that
explains the optical and X-ray data in region VI, assuming $t_{dec}\approx
3$\,d. As evident from Figure~\ref{fig:lclate}, because in a wind environment the optical
(and X-ray) light curves follow a rather flat temporal behavior, the model
overpredicts the optical observations at $t<t_{dec}$. Note also that an earlier
deceleration time would make this worse. We thus conclude that the peak
observed around 3\,d is more easily explained with the off-axis jet hypothesis
(ii): in regions V and VI, we are observing emission from a jet (hereafter
referred to as the late-time jet) which starts entering our line of sight (and
dominating the afterglow emission) in region V. This also implies (as we
discuss later on) that the emission observed in regions II-III-IV is likely
associated with a second jet (hereafter referred to as the early-time jet),
thus favoring a double jet scenario for \grb\,. We note that a similar model
has been proposed for several GRBs, such as GRB\,030329, GRB\,120404A, and GRB\,080319B
\citep{Berger:2003aa,Guidorzi:2014aa,Racusin:2008lr}.

The behavior of the radio emission in region V deserves special attention. Extrapolating the late-time X-ray
and optical data to the radio band via a simple power-law, overpredicts our radio observations by
2 orders of magnitude. Thus, if the radio peak we observe at 3\,d is dominated by FS emission from the 
late-time jet, then a spectral 
break between the optical and the radio bands is required. 
This constrains the location of $\nu_{m,f}$ at $3$\,d so that:
\begin{eqnarray}
\label{eq:numconstr}
F_{14.5 \rm{GHz}} ({\rm3\,d}) = F_{\rm Opt} ({\rm3\,d}) \times 
 (\nu_{m,f}/\nu_{\rm Opt})^{-\beta_{\rm OX}} \\
 \nonumber
(\nu_{\rm radio}/\nu_{m,f})^{1/3} < 0.46\,{\rm mJy} 
\end{eqnarray}
which implies $\nu_{m,f} (3\,\rm{d})>1.2\times 10^{12}$\,Hz and $F_{\nu_{m,f}}<2.0$\,mJy.

Moreover, if we assume the FS is solely responsible for the radio emission, the
low frequency observations at 11 and 16 days need to be consistent with
$\nu_{a,f}(\propto t^{-3/5})$ passing through the bands (also with the $21$ d SED
where $\nu_{a,f}(21 \,{\rm d})<3 $ GHz).  Accepting these constraints (see also Table~\ref{tab:specfit}), we can
find the FS self-absorption frequency at 3 d: $\nu_{a,f}({\rm
3\,d})=\nu_{a,f}({\rm 11\,d})(3/11)^{-3/5} \approx 20$\,GHz.  Using this
approximate value for $\nu_{a,f}$ and our conservative lower-limit for
$\nu_{m,f}$, we can now give a more quantitative estimate of the FS
contribution to the radio flux at 3\,d. We have $F_{\rm 14.5 GHz}({\rm
3\,d})=F_{\nu_{m,f}} (\nu_{a,f}/\nu_{m,f})^{1/3} (14.5\,{\rm
GHz}/\nu_{a,f})^2\lesssim 0.2$\,mJy. Because the measured flux is $F_{14.5\,\rm
GHz}(3\,\rm{d})\approx 0.46$\,mJy, this means that \textit{at 3\,d post burst,
at least 50\%} of the 14.5\,GHz flux is provided by a component \textit{other}
that the FS of the late-time jet. We suggest that this component is the RS
emission from such jet.  We also note that, in fact, if the 15\,GHz emission at
3\,d was dominated by FS emission at $\nu_{radio}<\nu_{a,f}$, we would expect
the emission at $t\gtrsim 3$\,d to \textit{rise} with time as $t^1$ until
$\nu_{radio}\approx \nu_{a,f}$, and then show a flat behavior ($t^0$) until
$\nu_{radio}\approx \nu_{m,f}$. This is not what we observe at 15\,GHz (see
Figure\,\ref{fig:lclaterad}).

\begin{figure*}[t!]
\epsscale{1.0}
\includegraphics[scale=0.60,angle=0]{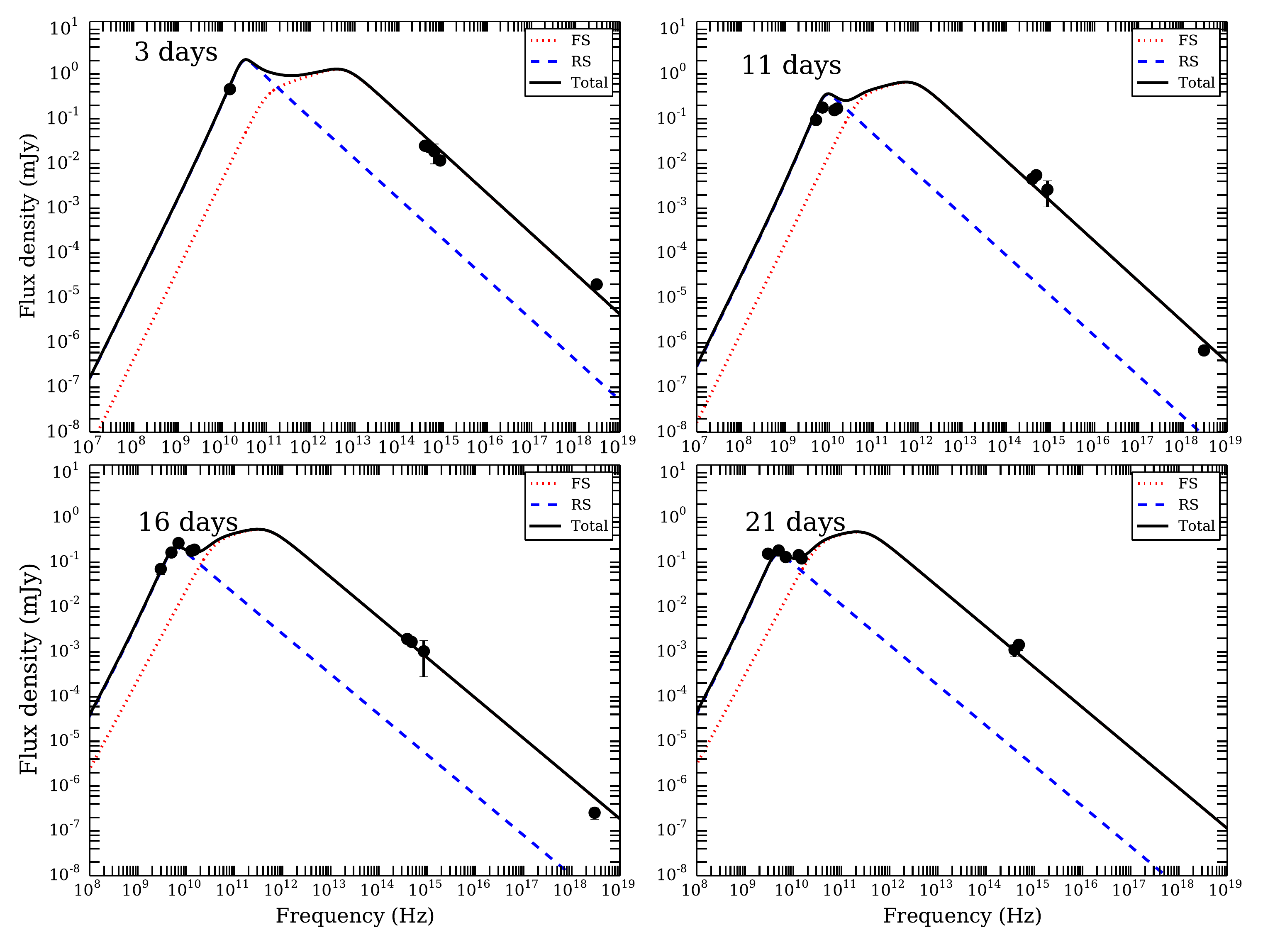}
\caption{\footnotesize{SEDs at 3, 11, 16 and 21 days after the burst: the FS (dashed red), RS (dashed blue) and FS+RS (solid black) 
contributions to the emission are shown together with the observed radio, optical and \xray\ data. The broken power law spectra for both FS and RS represent the standard synchrotron emission expected from a shocked electron distribution (Section \ref{sec:discussion}).} }
\label{fig:SEDs}
\end{figure*}

We model the radio-to-\xray\ in a scenario where the optical and X-ray
emission are FS-dominated and the 14.5\,GHz is RS-dominated.
Therefore we can  model the SED of \grb\, at 3\,d in the same context 
(Figure~ \ref{fig:SEDs}).In our model presented in Figure \ref{fig:lclaterad}, 
we are assuming a deceleration time of 3\,d for the
late-time jet and we are not attempting modeling the rise before 3\,d (since
for a jet entering our line of sight, one could have a large range of temporal
indices; see e.g. \citealt{Eichler:2006aa}). 

We finally note that alternative explanations for the $3$\,d peak, such as 
the passing of a characteristic frequency in band, can be excluded. Indeed, the passing of a characteristic 
frequency in optical or X-rays would imply a chromatic peak time and a spectral evolution across the peak \citep[as seen in other cases;][]{Guidorzi:2014aa}. The optical and \xray\ spectral indices of \grb\  before and after the 3\,d peak are consistent with no spectral evolution (see Table~\ref{tab:specfit}), within the (large) errors, while we do not have spectral information from the radio data around 3\,d. 

\subsubsection{Physical parameters}
\label{sec:phys}
In order to calculate the physical parameters for this burst, we proceed in the
following way: we identify the characteristic frequencies ($\nu_a$, $\nu_m$ and
$\nu_c$ for FS and RS) which determine the spectral and temporal evolution of
the afterglow. We
construct a model using these characteristic frequencies and compare it to our
observations. If we find a satisfactory agreement between model and
observations, in the next step we solve for the physical parameters that drive
the characteristic frequencies. 

First, we study the case where the characteristic frequencies have the
following ordering: $\nu_{a,f}<\nu_{m,f}<\nu_{c,f}$ and
$\nu_{a,r}<\nu_{m,r}<\nu_{c,r}$. We can put a constraint on  $\nu_{m,f}$ by
requiring the peak flux to lie on the extrapolation of the optical-to-X-ray
spectrum below the optical range (e.g. see Equation \ref{eq:numconstr} in  Section \ref{sec:3day}),
$\nu_{c,f}\gtrsim 2.4\times 10^{17} $ Hz and $\nu_{a,f}$ will be unconstrained,
because the RS dominates the flux at $\nu_{a,f}$. Similarly in the case of the
RS, only $\nu_{m,r}$ can be constrained.  We consider the characteristic
frequencies $\nu_{m,f}$ and $\nu_{c,f}$ at 3\,d as free parameters as well as
the total kinetic energy, $E$.  Finally, we use the expressions of $\nu_{m,f}$,
$\nu_{c,f}$ and $F_{\nu_m}$ \citep[e.g.  from][]{Granot:2002aa} to determine the efficiencies
$\epsilon_{B,f}$, $\epsilon_e$ and $A_\star$.  From the expression of $t_{\rm
dec}$ which we equate to $3\,d$ we can derive $\Gamma_0$.  Using the relations
between RS and FS characteristic quantities ($\nu_{m,r}\approx0.31~\Gamma_0
R_B^{1/2} \nu_{m,f}$, $\nu_{c,r}\approx R_B^{-3/2}\nu_{c,f}$ and
$F_{\nu_{m,r}}\approx 1.2~\Gamma_0 R_B^{1/2}F_{\nu_{m,f}}$, where
$R_B=\epsilon_{B,r}/\epsilon_{B,f}$ e.g.  \citet{Perley:2014aa}) we can
determine the RS quantities. Here, $R_B$ is also a free parameter.  We solve
the equations for the physical parameters by varying the free parameters ($E$,
$\nu_{m,f}$, $\nu_{c,f}$ and $R_B$) through the allowed parameter space (or a
sufficiently large range in the case of $E$ and $R_B$.), we find that either
the $\epsilon_e<1$ and $\epsilon_B<1$ or the $\nu_{a,r}>10 $ GHz condition
cannot be satisfied at the same time. 
Violating these conditions makes the solution non-physical, and in particular the latter
 is important in order for the RS to provide the
necessary 15\,GHz flux observed at 3\,d. We thus conclude that this ordering of
the frequencies cannot adequately reproduce the observations.

Next, we assume the RS peak is located at $\nu_{a,r}$, in other words the order
of frequencies in the RS is $\nu_{m,r}<\nu_{a,r}<\nu_{c,r}$. We proceed
similarly to the previous case: we set up the equations from the expressions of
$\nu_{m,f}$, $\nu_{c,f}$, $F_{\nu_{m,f}}$ and $T_{\rm dec}$. Additionally, we
consider the expression for $\nu_{a,r}=5.8\times10^{11} {\rm Hz}~ (1+z)^{-1}
(\epsilon_{e,-0.5} 3 (p-2)/(p-1))^{6/13} \epsilon_{B,r,-1}^{9/26} E_{54}^{-1}
\Gamma_{0,1.5}^{-4} A_{\star,-1}^{43/26}$ \citep{Zou:2005aa}.

We obtain a physically meaningful solution to the set of equations with the
following parameters: $\epsilon_e=0.405$, $\epsilon_{B,f}=0.023$,
$\epsilon_{B,r}=4.1\times10^{-3}$, $A_\star=0.05$ and $\Gamma_0=27.2$.    We also find a total kinetic energy of
$E=10^{54}$ erg,  a factor of $\sim10$ larger than the energy emitted in
gamma-rays and a value of $p=2.8$, which is within $2\sigma$ from the value
obtained independently from the late time lightcurve.
With these parameters, in addition to the model lightcurves, we construct spectral energy distributions for observations after 3 days and show that they provide an adequate description of the data (see Figure~\ref{fig:SEDs}).

\subsection{Regions II-IV}
One of the striking features of GRB\,141121A is the approximately constant
optical flux observed at early times in Region III, with hints of constant flux
as early as region I ($\approx 0.01$\,d; GROND and UVOT detections). In region
IV, before the rebrightening observed in region V, a decaying optical emission
is also observed: a second jet component, whose emission dominates at
early-times, could explain this unusual behavior. For example, some of the
light curves of a two-component jet observed slightly off-axis in Figure 4 of
\citealt{Huang:2004aa}, look qualitatively very similar to the optical light
curve of \grb.  We finally note that a two-component jet model with
contribution from a RS was also invoked by \citealt{van-der-Horst:2014aa} in
the case of GRB\,130427A.

\section{Conclusions}
\label{sec:conclusions} 
We have presented our multi-wavelength observing campaign of \grb\,, which was discovered by the \swift\ satellite and  observed starting a few hours after the explosion and continuing over the following month. The long duration of this burst places it in the class of UL-GRBs, providing one of the best cases to test the contribution of the RS
and its evolution in relation with the FS. Our extensive radio campaign, in combination with the identification
of an achromatic peak at $\approx3$\,d, enabled us to demonstrate that the RS is contributing at least 50\% of the observed flux, as well as that the complex optical light curve of this burst likely requires a two-component jet model. \grb\ is expanding in a wind-like environment, whose density appears to have an average value when compared to the distribution of values observed for other GRBs. 

The case of \grb\, shows the importance of combining rapid-response facilities (like RATIR) with \swift\ as well as with radio observations at various frequencies, overall constraining the temporal behavior of the GRB afterglow over $\sim10$ orders of magnitude in frequency. UL-GRBs are among the best transient objects for which 
we can test central engine theories and emission mechanisms, and future planned missions like SVOM (which covers from hard \xray\ to optical), XTIDE (designed to observe the transient \xray\ sky) or the ISS-Lobster concept will enable great steps forward in our understanding of such phenomena.


\acknowledgments
This research was supported by the NASA Postdoctoral Program at the Goddard Space Flight Center, 
administered by Oak Ridge Associated Universities through a contract with NASA.
AC thanks the PI of the Keck observations (Christian Ott) and the observers (Maryam Modjaz and David Fierroz) for 
donating some of their precious time to observe \grb.
A.Corsi acknowledges partial support from the NASA-\textit{Swift} GI program via grants 13-SWIFT13-0030 and 14-SWIFT14-0024.
Partial support of OTKA NN 111016 grant (PV).  
SBC acknowledges support from the NASA Fermi grant  NNH13ZDA001N.
Partial support for DAP was provided by NASA through an award issued by JPL/Caltech.
This work made use
of data supplied by the UK Swift Science Data Centre at the University of
Leicester.
We thank the CARMA observers (in particular G. Keating) for executing our observations.
Some of the data presented herein were obtained at the W.M. Keck Observatory, which is operated as a scientific partnership among the California Institute of Technology, the University of California and the National Aeronautics and Space Administration. The Observatory was made possible by the generous financial support of the W.M. Keck Foundation.
We thank the RATIR project team and the staff of the Observatorio Astronómico Nacional on Sierra San Pedro Mártir. RATIR is a collaboration between the University of California, the Universidad Nacional Autonóma de México, NASA Goddard Space Flight Center, and Arizona State University, benefiting from the loan of an H2RG detector and hardware and software support from Teledyne Scientific and Imaging. RATIR, the automation of the Harold L. Johnson Telescope of the Observatorio Astronómico Nacional on Sierra San Pedro Mártir, and the operation of both are funded through NASA grants NNX09AH71G, NNX09AT02G, NNX10AI27G, and NNX12AE66G, CONACyT grants INFR-2009-01-122785 and CB-2008-101958 , UNAM PAPIIT grant IN113810, IG100414, and UC MEXUS-CONACyT grant CN 09-283.
M. Modjaz is supported in parts by the NSF CAREER award AST-1352405 and
by the NSF award AST-1413260.
These results made use of Lowell Observatory?s Discovery Channel Telescope.
Lowell operates the DCT in partnership with Boston University, Northern Arizona University, the University
of Maryland, and the University of Toledo. Partial support of the DCT was provided by Discovery
Communications. LMI was built by Lowell Observatory using funds from the National Science Foundation
(AST-1005313).?


\clearpage

\input{tab1v2.tex}
\input{tab2Radio.tex}
\input{tab3specfitv2.tex}
\clearpage

\bibliographystyle{apj}
\bibliography{141121A,bibthesis}

\end{document}

%% file: defs.txt
\def\CIVdblt{{\rm C~}\kern 0.1em{\sc iv}~$\lambda\lambda 1548, 1550$}
\def\MgIIdblt{{\rm Mg~}\kern 0.1em{\sc ii}~$\lambda\lambda 2796, 2803$}
\def\NVdblt{{\rm N}\kern 0.1em{\sc v}~$\lambda\lambda 1238, 1242$}  
\def\OVIdblt{{\rm O}\kern 0.1em{\sc vi}~$\lambda\lambda 1031, 1037$}
\def\SiIVdblt{{\rm Si~}\kern 0.1em{\sc iv}~$\lambda\lambda1394, 1403$}
\def\AlIIIdblt{{\rm Al~}\kern 0.1em{\sc iii}~$\lambda\lambda1855,1863$}
\def\FeIIdblt{{\rm Fe~}\kern 0.1em{\sc ii}~$\lambda\lambda 2383, 2600$}

\def\AlII{\hbox{{\rm Al~}\kern 0.1em{\sc ii}}}
\def\AlIII{\hbox{{\rm Al~}\kern 0.1em{\sc iii}}}
\def\CaI{\hbox{{\rm Ca}\kern 0.1em{\sc i}}}
\def\CaII{\hbox{{\rm Ca}\kern 0.1em{\sc ii}}}

\def\CrII{\hbox{{\rm Cr~}\kern 0.1em{\sc ii}}}
\def\CI{\hbox{{\rm C~}\kern 0.1em{\sc i}}}
\def\CII{\hbox{{\rm C~}\kern 0.1em{\sc ii}}}
\def\CIII{\hbox{{\rm C~}\kern 0.1em{\sc iii}}}
\def\CIV{\hbox{{\rm C~}\kern 0.1em{\sc iv}}}
\def\CV{\hbox{{\rm C}\kern 0.1em{\sc v}}}

\def\H{\hbox{{\rm H}}}

\def\H2{\hbox{{\rm H$_2$}}}

\def\HII{\hbox{{\rm H}\kern 0.1em{\sc ii}}}

\def\Lya{\hbox{{\rm Ly}\kern 0.1em$\alpha$}}

\def\Lyb{\hbox{{\rm Ly}\kern 0.1em$\beta$}}
\def\Lyg{\hbox{{\rm Ly}\kern 0.1em$\gamma$}}
\def\Lyfive{\hbox{{\rm Ly}\kern 0.1em$5$}}
\def\Lysix{\hbox{{\rm Ly}\kern 0.1em$6$}}
\def\Lyseven{\hbox{{\rm Ly}\kern 0.1em$7$}}
\def\Lyeight{\hbox{{\rm Ly}\kern 0.1em$8$}}
\def\Lynine{\hbox{{\rm Ly}\kern 0.1em$9$}}
\def\Lyten{\hbox{{\rm Ly}\kern 0.1em$10$}}
\def\HeI{\hbox{{\rm He}\kern 0.1em{\sc i}}}
\def\HeII{\hbox{{\rm He}\kern 0.1em{\sc ii}}}
\def\FeI{\hbox{{\rm Fe~}\kern 0.1em{\sc i}}}
\def\FeII{\hbox{{\rm Fe~}\kern 0.1em{\sc ii}}}
\def\FeIII{\hbox{{\rm Fe~}\kern 0.1em{\sc iii}}}
\def\MnII{\hbox{{\rm Mn}\kern 0.1em{\sc ii}}}
\def\MgI{\hbox{{\rm Mg~}\kern 0.1em{\sc i}}}
\def\MgII{\hbox{{\rm Mg~}\kern 0.1em{\sc ii}}}
\def\MgIII{\hbox{{\rm Mg~}\kern 0.1em{\sc iii}}}
\def\MgIV{\hbox{{\rm Mg~}\kern 0.1em{\sc iv}}}
\def\NaI{\hbox{{\rm Na}\kern 0.1em{\sc i}}}
\def\NV{\hbox{{\rm N}\kern 0.1em{\sc v}}}
\def\NII{\hbox{{\rm N}\kern 0.1em{\sc ii}}}
\def\NIII{\hbox{{\rm N}\kern 0.1em{\sc iii}}}
\def\NiI{\hbox{{\rm Ni~}\kern 0.1em{\sc i}}}
\def\NiII{\hbox{{\rm Ni~}\kern 0.1em{\sc ii}}}
\def\OVI{\hbox{{\rm O}\kern 0.1em{\sc vi}}}
\def\OI{\hbox{{\rm O}\kern 0.1em{\sc i}}}
\def\OII{\hbox{[{\rm O}\kern 0.1em{\sc ii}]}}
\def\SiII{\hbox{{\rm Si~}\kern 0.1em{\sc ii}}}
\def\SiIII{\hbox{{\rm Si~}\kern 0.1em{\sc iii}}}
\def\SiIV{\hbox{{\rm Si~}\kern 0.1em{\sc iv}}}
\def\SII{\hbox{{\rm S~}\kern 0.1em{\sc ii}}}
\def\SIII{\hbox{{\rm S~}\kern 0.1em{\sc iii}}}
\def\SIV{\hbox{{\rm S~}\kern 0.1em{\sc iv}}}
\def\TiII{\hbox{{\rm Ti}\kern 0.1em{\sc ii}}}
\def\ZnII{\hbox{{\rm Zn~}\kern 0.1em{\sc ii}}}

\def\swift{\emph{Swift \,}}

\def\xray{X-ray}
\def\swift{\emph{Swift}}

\newcommand{\grb}{\mbox{GRB\,090429B}}
\def\simlt{\mathrel{\hbox{\rlap{\hbox{\lower4pt\hbox{$\sim$}}}\hbox{$<$}}}}
\def\simgt{\mathrel{\hbox{\rlap{\hbox{\lower4pt\hbox{$\sim$}}}\hbox{$>$}}}}

\newcommand{\up}{\mbox{$u^{\prime}$}}
\newcommand{\gp}{\mbox{$g^{\prime}$}}
\newcommand{\rp}{\mbox{$r^{\prime}$}}
\newcommand{\ip}{\mbox{$i^{\prime}$}}
\newcommand{\zp}{\mbox{$z^{\prime}$}}



%% file: tab1v2.tex
\begin{deluxetable}{lcccc}
\tablewidth{0in}
\tablecaption{Log of Observations\label{tab:data}}
\tabletypesize{\footnotesize} 
\tablehead{
\colhead{$T-T_0$} & \colhead{Mag}&\colhead{Flux} &  \colhead{Band} & \colhead{Instrument} \\
	(days)&&($\mu$Jy)&&
}
\startdata
\multicolumn{5}{c}{\emph{\small RATIR}}\\
\cline{1-5}
\noalign{\smallskip}
0.197  & $ 19.53 \pm 0.08$ & $ 56.05 \pm 4.35$ & \rp & RATIR \\  
0.239  & $ 19.50 \pm 0.07$ & $ 57.62 \pm 3.62$ & \rp & RATIR \\  
0.282  & $ 19.55 \pm 0.07$ & $ 54.79 \pm 3.58$ & \rp & RATIR \\  
0.324  & $ 19.59 \pm 0.07$ & $ 53.15 \pm 3.64$ & \rp & RATIR \\  
0.361  & $ 19.69 \pm 0.07$ & $ 48.28 \pm 3.05$ & \rp & RATIR \\  
1.152  & $ 20.67 \pm 0.07$ & $ 19.67 \pm 1.25$ & \rp & RATIR \\  
1.208  & $ 20.73 \pm 0.07$ & $ 18.61 \pm 1.12$ & \rp & RATIR \\  
1.264  & $ 20.77 \pm 0.07$ & $ 17.84 \pm 1.07$ & \rp & RATIR \\  
1.321  & $ 20.80 \pm 0.06$ & $ 17.43 \pm 1.01$ & \rp & RATIR \\  
1.365  & $ 20.78 \pm 0.04$ & $ 17.77 \pm 0.73$ & \rp & RATIR \\  
2.186  & $ 20.59 \pm 0.09$ & $ 21.08 \pm 1.67$ & \rp & RATIR \\  
2.241  & $ 20.53 \pm 0.06$ & $ 22.33 \pm 1.24$ & \rp & RATIR \\  
2.287  & $ 20.48 \pm 0.06$ & $ 23.40 \pm 1.22$ & \rp & RATIR \\  
2.346  & $ 20.54 \pm 0.06$ & $ 22.04 \pm 1.21$ & \rp & RATIR \\  
2.382  & $ 20.55 \pm 0.04$ & $ 21.84 \pm 0.88$ & \rp & RATIR \\  
3.167  & $ 20.55 \pm 0.09$ & $ 21.90 \pm 1.91$ & \rp & RATIR \\  
3.223  & $ 20.61 \pm 0.12$ & $ 20.62 \pm 2.30$ & \rp & RATIR \\  
3.279  & $ 20.60 \pm 0.08$ & $ 20.93 \pm 1.63$ & \rp & RATIR \\  
3.336  & $ 20.62 \pm 0.07$ & $ 20.59 \pm 1.28$ & \rp & RATIR \\  
3.371  & $ 20.60 \pm 0.05$ & $ 20.83 \pm 0.91$ & \rp & RATIR \\  
4.151  & $ 20.86 \pm 0.12$ & $ 16.50 \pm 1.87$ & \rp & RATIR \\  
4.210  & $ 21.01 \pm 0.15$ & $ 14.35 \pm 2.04$ & \rp & RATIR \\  
4.252  & $ 20.93 \pm 0.12$ & $ 15.49 \pm 1.76$ & \rp & RATIR \\  
4.302  & $ 20.83 \pm 0.10$ & $ 16.90 \pm 1.61$ & \rp & RATIR \\  
4.357  & $ 20.94 \pm 0.07$ & $ 15.28 \pm 1.01$ & \rp & RATIR \\  
5.187  & $ 21.12 \pm 0.09$ & $ 12.91 \pm 1.13$ & \rp & RATIR \\  
5.242  & $ 21.20 \pm 0.09$ & $ 11.99 \pm 1.04$ & \rp & RATIR \\  
5.296  & $ 21.15 \pm 0.08$ & $ 12.59 \pm 0.92$ & \rp & RATIR \\  
5.351  & $ 21.22 \pm 0.06$ & $ 11.76 \pm 0.68$ & \rp & RATIR \\  
6.182  & $ 21.39 \pm 0.14$ & $ 10.07 \pm 1.28$ & \rp & RATIR \\  
6.237  & $ 21.47 \pm 0.11$ & $ 9.35 \pm 0.98$ & \rp & RATIR \\  
6.292  & $ 21.56 \pm 0.12$ & $ 8.61 \pm 0.97$ & \rp & RATIR \\  
6.350  & $ 21.61 \pm 0.14$ & $ 8.22 \pm 1.07$ & \rp & RATIR \\  
6.385  & $ 21.58 \pm 0.10$ & $ 8.47 \pm 0.77$ & \rp & RATIR \\  
7.172  & $ 21.67 \pm 0.14$ & $ 7.79 \pm 1.03$ & \rp & RATIR \\  
7.228  & $ 21.74 \pm 0.13$ & $ 7.30 \pm 0.90$ & \rp & RATIR \\  
7.283  & $ 21.93 \pm 0.17$ & $ 6.16 \pm 0.99$ & \rp & RATIR \\  
7.339  & $ 21.97 \pm 0.20$ & $ 5.94 \pm 1.12$ & \rp & RATIR \\  
7.379  & $ 22.05 \pm 0.14$ & $ 5.48 \pm 0.68$ & \rp & RATIR \\  
8.167  & $ 21.91 \pm 0.16$ & $ 6.26 \pm 0.93$ & \rp & RATIR \\  
8.223  & $ 22.09 \pm 0.20$ & $ 5.29 \pm 0.97$ & \rp & RATIR \\  
8.279  & $ 22.02 \pm 0.17$ & $ 5.63 \pm 0.87$ & \rp & RATIR \\  
8.336  & $ 22.25 \pm 0.19$ & $ 4.58 \pm 0.80$ & \rp & RATIR \\  
8.371  & $ 22.30 \pm 0.14$ & $ 4.37 \pm 0.57$ & \rp & RATIR \\  
9.161  & $ 22.11 \pm 0.24$ & $ 5.20 \pm 1.16$ & \rp & RATIR \\  
9.236  & $ 22.28 \pm 0.24$ & $ 4.45 \pm 0.98$ & \rp & RATIR \\  
9.298  & $ 22.27 \pm 0.23$ & $ 4.49 \pm 0.95$ & \rp & RATIR \\  
9.362  & $ 21.98 \pm 0.13$ & $ 5.86 \pm 0.68$ & \rp & RATIR \\  
10.158  & $ 22.05 \pm 0.20$ & $ 5.50 \pm 1.02$ & \rp & RATIR \\  
16.245  & $ 23.33 \pm 0.23$ & $ 1.68 \pm 0.36$ & \rp & RATIR \\  
21.261  & $ 23.50 \pm 0.23$ & $ 1.44 \pm 0.31$ & \rp & RATIR \\  
\cline{1-5}
\hline\noalign{\smallskip}
0.198  & $ 19.31 \pm 0.08$ & $ 68.60 \pm 4.75$ & \ip & RATIR \\  
0.239  & $ 19.33 \pm 0.07$ & $ 67.05 \pm 4.02$ & \ip & RATIR \\  
0.282  & $ 19.38 \pm 0.07$ & $ 64.41 \pm 3.97$ & \ip & RATIR \\  
0.324  & $ 19.43 \pm 0.07$ & $ 61.54 \pm 3.93$ & \ip & RATIR \\  
0.361  & $ 19.48 \pm 0.06$ & $ 58.50 \pm 3.21$ & \ip & RATIR \\  
1.152  & $ 20.56 \pm 0.07$ & $ 21.77 \pm 1.40$ & \ip & RATIR \\  
1.208  & $ 20.60 \pm 0.06$ & $ 20.82 \pm 1.23$ & \ip & RATIR \\  
1.264  & $ 20.67 \pm 0.06$ & $ 19.63 \pm 1.14$ & \ip & RATIR \\  
1.321  & $ 20.68 \pm 0.06$ & $ 19.39 \pm 1.15$ & \ip & RATIR \\  
1.365  & $ 20.72 \pm 0.05$ & $ 18.72 \pm 0.78$ & \ip & RATIR \\  
2.186  & $ 20.46 \pm 0.09$ & $ 23.72 \pm 1.98$ & \ip & RATIR \\  
2.241  & $ 20.37 \pm 0.06$ & $ 25.72 \pm 1.46$ & \ip & RATIR \\  
2.287  & $ 20.40 \pm 0.06$ & $ 25.02 \pm 1.40$ & \ip & RATIR \\  
2.346  & $ 20.36 \pm 0.06$ & $ 26.08 \pm 1.41$ & \ip & RATIR \\  
2.382  & $ 20.39 \pm 0.04$ & $ 25.43 \pm 1.03$ & \ip & RATIR \\  
3.184  & $ 20.37 \pm 0.11$ & $ 25.83 \pm 2.66$ & \ip & RATIR \\  
3.243  & $ 20.46 \pm 0.12$ & $ 23.72 \pm 2.61$ & \ip & RATIR \\  
3.298  & $ 20.37 \pm 0.08$ & $ 25.76 \pm 1.81$ & \ip & RATIR \\  
3.353  & $ 20.52 \pm 0.05$ & $ 22.54 \pm 1.03$ & \ip & RATIR \\  
4.151  & $ 20.71 \pm 0.12$ & $ 18.95 \pm 2.10$ & \ip & RATIR \\  
4.210  & $ 20.82 \pm 0.14$ & $ 17.07 \pm 2.14$ & \ip & RATIR \\  
4.252  & $ 20.66 \pm 0.10$ & $ 19.81 \pm 1.92$ & \ip & RATIR \\  
4.302  & $ 20.75 \pm 0.11$ & $ 18.24 \pm 1.78$ & \ip & RATIR \\  
4.357  & $ 20.73 \pm 0.07$ & $ 18.57 \pm 1.14$ & \ip & RATIR \\  
5.187  & $ 20.93 \pm 0.09$ & $ 15.39 \pm 1.23$ & \ip & RATIR \\  
5.242  & $ 20.95 \pm 0.08$ & $ 15.11 \pm 1.17$ & \ip & RATIR \\  
5.296  & $ 21.06 \pm 0.08$ & $ 13.73 \pm 1.04$ & \ip & RATIR \\  
5.351  & $ 20.94 \pm 0.06$ & $ 15.34 \pm 0.80$ & \ip & RATIR \\  
6.182  & $ 21.36 \pm 0.14$ & $ 10.37 \pm 1.37$ & \ip & RATIR \\  
6.237  & $ 21.41 \pm 0.11$ & $ 9.87 \pm 0.96$ & \ip & RATIR \\  
6.292  & $ 21.35 \pm 0.11$ & $ 10.49 \pm 1.04$ & \ip & RATIR \\  
6.350  & $ 21.40 \pm 0.13$ & $ 9.99 \pm 1.16$ & \ip & RATIR \\  
6.275  & $ 21.31 \pm 0.04$ & $ 10.82 \pm 0.44$ & \ip & RATIR \\  
7.172  & $ 21.50 \pm 0.13$ & $ 9.14 \pm 1.07$ & \ip & RATIR \\  
7.228  & $ 21.58 \pm 0.12$ & $ 8.49 \pm 0.97$ & \ip & RATIR \\  
7.283  & $ 21.49 \pm 0.12$ & $ 9.21 \pm 0.98$ & \ip & RATIR \\  
7.339  & $ 21.64 \pm 0.15$ & $ 8.03 \pm 1.09$ & \ip & RATIR \\  
7.379  & $ 21.56 \pm 0.09$ & $ 8.60 \pm 0.71$ & \ip & RATIR \\  
8.167  & $ 21.91 \pm 0.15$ & $ 6.28 \pm 0.89$ & \ip & RATIR \\  
8.223  & $ 21.85 \pm 0.14$ & $ 6.62 \pm 0.86$ & \ip & RATIR \\  
8.279  & $ 21.82 \pm 0.13$ & $ 6.78 \pm 0.82$ & \ip & RATIR \\  
8.336  & $ 21.95 \pm 0.14$ & $ 6.05 \pm 0.80$ & \ip & RATIR \\  
8.371  & $ 21.95 \pm 0.11$ & $ 6.04 \pm 0.58$ & \ip & RATIR \\  
9.161  & $ 22.00 \pm 0.23$ & $ 5.74 \pm 1.21$ & \ip & RATIR \\  
9.216  & $ 22.03 \pm 0.19$ & $ 5.62 \pm 0.99$ & \ip & RATIR \\  
9.271  & $ 22.12 \pm 0.18$ & $ 5.17 \pm 0.88$ & \ip & RATIR \\  
9.326  & $ 22.18 \pm 0.20$ & $ 4.88 \pm 0.89$ & \ip & RATIR \\  
9.371  & $ 21.91 \pm 0.12$ & $ 6.26 \pm 0.70$ & \ip & RATIR \\  
10.158  & $ 22.26 \pm 0.22$ & $ 4.54 \pm 0.92$ & \ip & RATIR \\  
16.245  & $ 23.34 \pm 0.22$ & $ 1.68 \pm 0.35$ & \ip & RATIR \\  
21.261  & $ 23.79 \pm 0.32$ & $ 1.11 \pm 0.33$ & \ip & RATIR \\  
\hline\noalign{\smallskip}
\multicolumn{5}{c}{\emph{\small Palomar P60}}\\
\cline{1-5}
\hline\noalign{\smallskip}
1.270  & $ 20.70 \pm 0.03$ & $ 19.02 \pm 0.51$ & \rp & Palomar-P60 \\  
2.250  & $ 20.57 \pm 0.08$ & $ 21.40 \pm 1.50$ & \rp & Palomar-P60 \\  
3.250  & $ 20.80 \pm 0.09$ & $ 17.44 \pm 1.51$ & \rp & Palomar-P60 \\  
4.170  & $ 20.82 \pm 0.06$ & $ 17.05 \pm 0.93$ & \rp & Palomar-P60 \\  
5.300  & $ 21.16 \pm 0.07$ & $ 12.49 \pm 0.83$ & \rp & Palomar-P60 \\  
6.160  & $ 21.52 \pm 0.10$ & $ 8.97 \pm 0.78$ & \rp & Palomar-P60 \\  
1.300  & $ 20.68 \pm 0.04$ & $ 18.55 \pm 0.62$ & \ip & Palomar-P60 \\  
2.240  & $ 20.37 \pm 0.07$ & $ 20.87 \pm 1.37$ & \ip & Palomar-P60 \\  
3.240  & $ 20.45 \pm 0.09$ & $ 17.01 \pm 1.35$ & \ip & Palomar-P60 \\  
4.190  & $ 20.59 \pm 0.05$ & $ 16.63 \pm 0.74$ & \ip & Palomar-P60 \\  
5.290  & $ 20.91 \pm 0.07$ & $ 12.18 \pm 0.75$ & \ip & Palomar-P60 \\  
6.180  & $ 21.17 \pm 0.08$ & $ 8.75 \pm 0.68$ & \ip & Palomar-P60 \\  
2.220  & $ 20.72 \pm 0.05$ & $ 18.67 \pm 0.89$ & \gp & Palomar-P60 \\  
3.220  & $ 20.81 \pm 0.06$ & $ 17.17 \pm 0.92$ & \gp & Palomar-P60 \\  
4.220  & $ 21.17 \pm 0.05$ & $ 12.41 \pm 0.55$ & \gp & Palomar-P60 \\  
\hline\noalign{\smallskip}
\multicolumn{5}{c}{\emph{\small Discovery Channel Telescope}}\\
\cline{1-5}
\hline\noalign{\smallskip}
16.098  & $ 23.34 \pm 0.14$ & $ 1.67 \pm 0.21$ & \rp & DCT \\  
16.098  & $ 23.18 \pm 0.12$ & $ 1.94 \pm 0.21$ &  \ip  & DCT \\  
\hline\noalign{\smallskip}
\multicolumn{5}{c}{\emph{\small UVOT}}\\
\cline{1-5}
\hline\noalign{\smallskip}
0.075  & $ 20.21 \pm 0.29$ & $ 29.79 \pm 7.85$ & \emph{u} & UVOT \\  
0.335  & $ 20.12 \pm 0.11$ & $ 32.48 \pm 3.38$ & \emph{u} & UVOT \\  
0.695  & $ 20.86 \pm 0.09$ & $ 16.52 \pm 1.42$ & \emph{u} & UVOT \\  
0.797  & $ 20.68 \pm 0.11$ & $ 19.39 \pm 1.91$ & \emph{u} & UVOT \\  
1.768  & $ 21.19 \pm 0.14$ & $ 12.12 \pm 1.62$ & \emph{u} & UVOT \\  
2.624  & $ 21.23 \pm 0.18$ & $ 11.70 \pm 1.97$ & \emph{u} & UVOT \\  
3.541  & $ 21.28 \pm 0.13$ & $ 11.13 \pm 1.34$ & \emph{u} & UVOT \\  
4.607  & $ 21.78 \pm 0.22$ & $ 7.02 \pm 1.40$ & \emph{u} & UVOT \\  
6.705  & $ 22.42 \pm 0.20$ & $ 3.90 \pm 0.72$ & \emph{u} & UVOT \\  
8.683  & $ 22.73 \pm 0.36$ & $ 2.95 \pm 0.98$ & \emph{u} & UVOT \\  
10.448  & $ 22.86 \pm 0.64$ & $ 2.61 \pm 1.53$ & \emph{u} & UVOT \\  
14.266  & $ 23.75 \pm 0.71$ & $ 1.15 \pm 0.75$ & \emph{u} & UVOT \\  
0.079  & $ 19.45 \pm 0.31$ & $ 60.20 \pm 17.24$ & \emph{b} & UVOT \\  
0.215  & $ 20.08 \pm 0.69$ & $ 33.67 \pm 21.49$ & \emph{b} & UVOT \\  
0.681  & $ 20.26 \pm 0.22$ & $ 28.66 \pm 5.73$ & \emph{b} & UVOT \\  
1.777  & $ 20.80 \pm 0.33$ & $ 17.33 \pm 5.32$ & \emph{b} & UVOT \\  
0.192  & $ 20.21 \pm 0.40$ & $ 29.81 \pm 10.85$ & \emph{v} & UVOT \\  
0.055  & $ 20.72 \pm 0.56$ & $ 18.73 \pm 9.66$ & \emph{uvw1} & UVOT \\  
0.265  & $ 20.80 \pm 0.30$ & $ 17.40 \pm 4.77$ & \emph{uvw1} & UVOT \\  
0.326  & $ 20.88 \pm 0.14$ & $ 16.17 \pm 2.03$ & \emph{uvw1} & UVOT \\  
1.764  & $ 22.22 \pm 0.27$ & $ 4.70 \pm 1.18$ & \emph{uvw1} & UVOT \\  
2.593  & $ 21.91 \pm 0.14$ & $ 6.26 \pm 0.82$ & \emph{uvw1} & UVOT \\  
4.884  & $ 22.33 \pm 0.18$ & $ 4.23 \pm 0.70$ & \emph{uvw1} & UVOT \\  
5.424  & $ 22.49 \pm 0.24$ & $ 3.67 \pm 0.83$ & \emph{uvw1} & UVOT \\  
7.812  & $ 23.24 \pm 0.38$ & $ 1.84 \pm 0.65$ & \emph{uvw1} & UVOT \\  
9.751  & $ 22.86 \pm 0.27$ & $ 2.60 \pm 0.64$ & \emph{uvw1} & UVOT \\  
13.531  & $ 24.40 \pm 1.78$ & $ 0.63 \pm 1.04$ & \emph{uvw1} & UVOT \\  
\hline\noalign{\smallskip}
\multicolumn{5}{c}{\emph{\small GCN}}\\
\cline{1-5}
\hline\noalign{\smallskip}
0.505  & $ 20.13 \pm 0.08$ & $ 32.06 \pm 2.36$ & \up & Keck-LRIS \\  
0.408  & $ 19.81 \pm 0.01$ & $ 43.33 \pm 0.40$ & \gp & Keck-LRIS \\  
0.505  & $ 19.99 \pm 0.02$ & $ 36.71 \pm 0.68$ & \gp & Keck-LRIS \\  
27.305  & $ 24.46 \pm 0.04$ & $ 0.60 \pm 0.02$ & \gp & Keck-LRIS \\ 
0.015  & $ 19.70 \pm 0.10$ & $ 48.09 \pm 4.43$ & \gp & GROND$^1$ \\  
0.302  & $ 19.52 \pm 0.06$ & $ 56.76 \pm 3.14$ &  \rp&LCO-FTN$^2$\\  
0.385  & $ 19.71 \pm 0.10$ & $ 47.21 \pm 4.35$ & \rp& MITSuME$^3$ \\ 
0.408  & $ 19.45 \pm 0.01$ & $ 59.98 \pm 0.55$ &  \rp&Keck-LRIS\\ 
0.505  & $ 19.68 \pm 0.02$ & $ 48.98 \pm 0.90$ & \rp & Keck-LRIS\\
1.819  & $ 20.82 \pm 0.03$ & $ 17.14 \pm 0.47$ & \rp & TSHAO$^4$ \\  
2.320  & $ 20.50 \pm 0.04$ & $ 22.80 \pm 0.84$ &  \rp &LCO-FTN$^2$ \\  
5.830  & $ 21.54 \pm 0.09$ & $ 8.83 \pm 0.73$ &  \rp &TSHAO$^6$ \\  
27.305  & $ 24.16 \pm 0.09$ & $ 0.79 \pm 0.07$ & \rp & Keck-LRIS \\
0.015  & $ 19.52 \pm 0.10$ & $ 56.39 \pm 5.19$ & \ip&  GROND$^1$ \\  
0.310  & $ 19.34 \pm 0.06$ & $ 66.56 \pm 3.68$ & \ip  & LCO-FTN$^2$ \\  
0.385  & $ 19.66 \pm 0.10$ & $ 49.57 \pm 4.57$ & \ip  & MITSuME$^3$ \\  
2.300  & $ 20.26 \pm 0.05$ & $ 28.53 \pm 1.31$ &\ip & LCO-FTN$^2$ \\  
27.305  & $ 23.88 \pm 0.07$ & $ 1.02 \pm 0.07$ &  \ip & Keck-LRIS\\

\enddata
\tablecomments{Magnitude presented are corrected for galactic extinction using  \cite{Schlafly:2011aa}}
\tablecomments{References: (1)\cite{Tanga:2014aa}; (2) \cite{Dichiara:2014aa}; \cite{Kurita:2014aa}; 
(4)\cite{Volnova:2014aa}; (5)\cite{Dichiara:2014ab}; (6)\cite{Mazaeva:2014aa} 
}
\end{deluxetable}

%% file: tab2Radio.tex
\begin{deluxetable}{lccc}
\tablewidth{0in}
\tablecaption{Log of Radio Observations\label{tab:radio}}
\tabletypesize{\footnotesize} 
\tablehead{
\colhead{$T-T_0$} & \colhead{Flux} &  \colhead{Band} & \colhead{Instrument} \\
	(days)&($\mu$Jy)&&
}
\startdata
0.536  & $ \lesssim1500$ & 93 GHz  & CARMA \\  
0.910  & $ 180 \pm 60$ & 15 GHz  & AMI-LA  \\  
1.308  & $ \lesssim1150$ & 93 GHz  & CARMA \\  
3.037  & $ 460 \pm 50$ & 15 GHz  & AMI-LA  \\  
5.433  & $ \lesssim1160$ & 93 GHz  & CARMA \\  
6.016  & $ 170 \pm 40$ & 4.9 GHz  & WSRT \\  
8.352  & $ 204 \pm 12$ & 13 GHz  & VLA \\  
8.352  & $ 222 \pm 13$ & 15 GHz  & VLA \\  
11.402  & $ 170 \pm 11$ & 15 GHz  & VLA \\  
11.402  & $ 94.0 \pm 9.8$ & 5 GHz  & VLA \\  
11.402  & $ 179 \pm 12$ & 7 GHz  & VLA \\  
11.402  & $ 156 \pm 10$ & 13 GHz  & VLA \\  
16.385  & $ 71 \pm 17$ & 3 GHz  & VLA \\  
16.385  & $ 166 \pm 15$ & 5 GHz  & VLA \\  
16.385  & $ 270 \pm 18$ & 7 GHz  & VLA \\  
16.385  & $ 181 \pm 13$ & 13 GHz  & VLA \\  
16.385  & $ 192 \pm 14$ & 15 GHz  & VLA \\  
21.350  & $ 156 \pm 21$ & 3 GHz  & VLA \\  
21.350  & $ 184 \pm 15$ & 5 GHz  & VLA \\  
21.350  & $ 131 \pm 15$ & 7 GHz  & VLA \\ 
21.350  & $ 144.9 \pm 9.4$ & 13 GHz  & VLA \\  
21.350  & $ 122.0 \pm 8.7$ & 15 GHz  & VLA \\  
24.329  & $ 126 \pm 17$ & 3 GHz  & VLA \\  
24.329  & $ 112\pm 11$ & 5 GHz  & VLA \\  
24.329  & $ 141 \pm 14$ & 7 GHz  & VLA \\  
24.329  & $ 140 \pm 9$ & 13 GHz  & VLA \\  
24.329  & $ 122 \pm 9$ & 15 GHz  & VLA \\  
28.323  & $ 120 \pm 17$ & 3 GHz  & VLA \\  
28.323  & $ 111 \pm 11$ & 5 GHz  & VLA \\  
28.323  & $ 118 \pm 12$ & 7 GHz  & VLA \\  
28.323  & $ 106.7 \pm 8.1$ & 13 GHz  & VLA \\  
28.323  & $ 92.0 \pm 7.8$ & 15 GHz  & VLA \\  
33.350  & $ 177 \pm 25$ & 3 GHz  & VLA \\
33.350  & $ 168 \pm 15$ & 5 GHz  & VLA \\
33.350  & $ 123 \pm 13$ & 7 GHz  & VLA \\
33.350  & $ 101.0 \pm 8.7$ & 13 GHz  & VLA \\ 
33.350  & $ 108 \pm 7$ & 15 GHz  & VLA \\
\enddata
\tablecomments{Radio observations obtained with the VLA and CARMA facilities. We also list some publicly available 
data obtained with the AMI-LA telescope \citep{Anderson:2014aa} and WSRT \citep{vanDerHorst:2014}.}
\end{deluxetable}

%% file: tab3specfitv2.tex
\begin{deluxetable}{lll}
\tablewidth{0in}
\tablecaption{Spectral analysis\label{tab:specfit}}
\tabletypesize{\footnotesize} 
\tablehead{
\colhead{Region} & \colhead{Temporal} &  \colhead{Spectral} \\
	   		&index  & index \\
}
\startdata
Region \rm{I}  &...&$\beta_{Opt}=0.74\pm 0.47$\\
		     &...&$\beta_{X}=0.55\pm0.04$\\
Region \rm{II}  &...&$\beta_{Opt}=3.3\pm0.17$\\
			  &...&$\beta_{X}=0.92\pm0.17$\\

Region \rm{III}  &$\alpha_{Opt}=0.15\pm 0.11$&$\beta_{Opt}=0.87\pm0.14$\\
			  &$\alpha_{X}=3.17\pm0.14$&$\beta_{X}=0.92\pm0.13$\\
Region \rm{IV}  &$\alpha_{Opt}=0.84\pm0.11$&$\beta_{Opt}=0.29\pm0.21$\\
			  &...&$\beta_{X}=0.83\pm0.21$\\

Region \rm{V}  &$\alpha_{Opt}^{rise}=-1.77\pm0.77$&$\beta_{Opt}^{rise}=0.49\pm0.18$\\
			& $\alpha_{Opt}^{decay}=1.84\pm 0.17$   &$\beta_{Opt}^{decay}=0.83\pm0.16$\\ 
			  &$\alpha_{X}^{rise}=-2.33\pm0.88$&$\beta_{X}=0.67\pm0.23$\\  
			& $\alpha_{X}^{decay}=2.86\pm0.21$ &\\
Region \rm{VI}  &$\alpha_{Opt}=1.65\pm0.40$&$\beta_{Opt}=0.84\pm 0.47$\\
  		&$\alpha_{X}=1.85\pm0.34$&$\beta_{X}=0.86\pm0.33$\\
		  &$\alpha_{3 GHz}=-0.87\pm0.32$&$\beta_{radio}(11\,{\rm d)}=-0.34\pm0.05$\\
		 & $\alpha_{5GHz}=-0.19\pm0.10$&$\beta_{radio}(16\,{\rm d)}=-0.16\pm0.06$\\
		  &$\alpha_{7GHz}=0.45\pm0.09$&$\beta_{radio}(21\,{\rm d)}=+0.18\pm0.07$\\
		  &$\alpha_{13GHz}=0.43\pm0.05$&$\beta_{radio}(25\,{\rm d)}=-0.06\pm0.07$\\
		  &$\alpha_{15GHz}=0.57\pm0.05$&$\beta_{radio}(28\,{\rm d)}=+0.12\pm0.06$\\
		  &...&$\beta_{radio}(33\,{\rm d)}=+0.37\pm0.08$\\
\enddata
\tablecomments{Temporal and spectral analysis results for the different regions. In regions I and II the temporal indices
in optical and \xray\ are not calculated because of lack of measurements (optical) or rapid variation within the same region (\xray). See the main text for more details.}
\end{deluxetable}